\documentstyle[prb,aps,12pt]{revtex}
\begin{document} 
  
\title{Normal state properties of an interacting large polaron gas}
\author{G.~De~Filippis$^*$, V.~Cataudella$^\dagger$, 
G.~Iadonisi$^\dagger$}
\address {$^*$Dipartimento di Scienze Fisiche, 
Universit\`a di Salerno I-84081 Baronissi (Salerno), Italy} 
\address{$^\dagger$Dipartimento di Scienze Fisiche, 
Universit\`a di Napoli I-80125 Napoli, Italy} 
\date {\today} 
\maketitle 
\begin {abstract} 
A simple approach to the many-polaron problem for both weak and intermediate 
electron-phonon coupling and valid for densities much smaller than those 
typical of metals is presented. Within the model the total energy, the 
collective modes and the single-particle properties are studied and compared 
with the available theories. It is shown the occurrence of a charge density 
wave instability in the intermediate coupling regime. 
\end {abstract} 
\pacs{PACS: 71.38 (Polarons)  } 

\newpage
\section {Introduction}

The formation of large polarons and bipolarons in polar materials due
to Fr\"ohlich interaction with longitudinal optical (LO) phonons has
been studied quite extensively since the pioneering work of Landau
\cite{landau}, Pekar\cite{pekar} 
and Fr\"ohlich\cite{frohlich} and now it can be considered a well
understood problem\cite{devreese1}. However, a large amount of work has 
been devoted to the simpler single polaron and bipolaron problems neglecting
the effect of the polaron-polaron interaction. These effects, on the other
hand, are expected to play an important role in heavily doped semiconductors
\cite{mahan} and in many doped perovskites including high $T_c$ 
superconductors\cite{devreese2,emin} that are both characterized by
large polar effects and densities much smaller than those typical of metals.
In particular, in the case of superconductors infrared absorption
measurements suggest the existence of polarons whose properties 
depend strongly on doping\cite{calvani}. 

Recently, the regime characterized by low charge carrier 
density and strong e-ph interaction 
has been analyzed. In this regime the formation of a polaron Wigner crystal 
state is favoured with respect to the metallic phase. 
The stability of such a state in an ionic dielectric has been studied by the 
path integral 
technique pointing out the competition between the dissociation of the 
polarons at the insulator-metal transition and the melting towards a polaron 
liquid\cite{quemerais}. 
Moreover  two different approaches are known in literature for the 
metallic phase of a large polaron system. 
A perturbative approach, mainly due
to Mahan \cite {mahan,mahan1}, 
makes use of the random phase approximation (R.P.A.) in order to obtain a 
retarded effective electron-electron (e-e) interaction due to the exchange of 
LO phonons \cite{note}.
A variational approach, due to Devreese\cite{devreese}, 
is based on an 
extension of the Lee, Low and Pines (LLP)\cite{lee} canonical transformation 
to the many body problem and is able to give the total energy of the system
in terms of the electronic static structure factor. However, both approaches
are satisfactory only when electron-phonon (e-ph) effects are weak. 

In order to study the many polaron effects in the weak and intermediate 
e-ph interaction ($\alpha<7$)
and for polaron densities typical of the metallic phase of the 
doped semiconductors and
perovskites we introduce a simple model in which the electron-phonon
and the electron-electron interactions are not taken into account
at the same level. First the hamiltonian of two electrons interacting with
each other through Coulomb repulsion and with longitudinal optical phonons 
is introduced and an effective self-consistent electron-electron potential, 
due to LO phonons exchange, is determined within a variational approach. 
Then the many body effects are taken into account considering many electrons 
interacting with each other through the obtained effective potential. 
Within the proposed model we
show that, for weak e-ph interaction, many of the known results
can be recovered, while,   
for intermediate e-ph interaction, we present evidences for a 
charge density wave instability as suggested by several authors
\cite{dicastro,kivelson}. 
 
The paper is organized in the following way. In Section II the model
is introduced and its validity is discussed. The perturbative and the 
variational approaches are also briefly reviewed. In Section III the total 
energy, the collective excitations
and the quasi-particle self-energy are calculated for weak e-ph 
coupling and are compared with the results obtained within the variational and 
perturbative approaches. In Sec. IV the region of intermediate 
e-ph interaction is analyzed 
and numerical results for the single-particle spectral weight function, the 
collective excitation spectrum and the renormalization coefficient are 
presented just above the appearance of the instability signaled by the 
complete softening of the collective mode.
\section {Theoretical framework}
In systems characterized by charge carriers coupled with a polar lattice 
one has to consider the electron-electron interaction and the 
electron-phonon interaction. The Hamiltonian, which can be used to 
describe these materials, has the following form\cite{mahan}:
\begin {equation}
H=H_{0}+H_{e-e}+H_{e-ph} 
\label{1r} 
\end {equation} 
where 
\[ 
H_{0}=\sum_{\vec{p}\sigma} E^{0}(k)c^{\dagger}_{\vec{p}\sigma}c_{\vec{p}
\sigma}+\sum_{\vec{q}}\hbar\omega_{l}a^{\dagger}_{\vec{q}}a_{\vec{q}} 
\] 
\[
H_{e-e}=\frac{1}{2V}\sum_{\stackrel{\vec{p}_1\vec{p}_2\vec{q}}
{\sigma_1\sigma_2}}
v^{\infty}_{q}c^{\dagger}_{\vec{p}_1+\vec{q}\sigma_1}c^{\dagger}_{\vec{p}_
2-\vec{q}\sigma_2}c_{\vec{p}_2\sigma_2}c_{\vec{p}_1\sigma_1} 
\] 
\[
H_{e-ph}=\sum_{\stackrel{\vec{p}_1\vec{p}_2\vec{q}}{\sigma}}
M_qc^{\dagger}_{\vec{p}+\vec{q}\sigma}c_{\vec{p}\sigma}\left(a_{\vec{q}}+
a^{\dagger}_{-\vec{q}} \right)
\]
\[
v^{\infty}_{q}=\frac{v_{q}}{\epsilon_{\infty}}=\frac{4\pi 
e^{2}}{\epsilon
_{\infty}q^{2}}~.
\]	
In Eq.(\ref{1r}) the first term represents the kinetic energy of the 
electrons with effective band mass $m$ and the energy of the free 
longitudinal optical phonons; the second term is the Hamiltonian of the 
electrons interacting with the Coulomb potential screened by the 
background high frequency dielectric constant $\epsilon_{\infty}$ and 
the last term describes the electron-phonon interaction whose strength, 
in the Fr\"ohlich scheme\cite{frohlich}, is given by: 
\begin {equation} 
M_{q}=-i\hbar\omega_{l}\frac{R^{1/2}_{p}}{q}\left(\frac{4\pi\alpha}{V} 
\right)^{1/2} 
\label{2r} 
\end {equation} 
where $\alpha$ is the Fr\"ohlich coupling constant, 
$R_{p}=\left(\hbar/\left(2m\omega_{l}\right)\right)^{1/2}$ is the polaron 
radius and $V$ is the volume of the system. 
In $H_{e-e}$ and in $H_{e-ph}$ 
the $q=0$ term is omitted in the sums: i.e., we 
suppose that the self-energy of an uniform positive charge is subtracted 
from Eq.(\ref{1r}).  
The main properties of the metallic phase of a system described by 
Eq.(\ref{1r}) have been discussed within two approaches, proposed by 
Devreese\cite{devreese} and Mahan\cite{mahan,mahan1} respectively. 
In order to make clearer the comparison with the model which we will 
discuss, we start with 
a brief review of the two approaches.

The former is a variational calculation of the ground state energy 
which makes use of a generalization of the LLP transformation
\cite{lee} for many polaron systems. The energy per particle of the 
polaron gas is written as: 
\begin {equation} 
E_T=E_{c}+E_{p} 
\label{3r} 
\end {equation} 
where $E_{c}$ is the contribution due to the coulomb interaction and 
$E_{p}$ represents the e-ph contribution 
screened by the electrons, which can be expressed in terms of the electronic 
static structure factor $S(q)$: 
\begin {equation} 
E_{p}=-\sum_{\vec{q}}\frac{|M_{q}|^{2}S(q)}
{\hbar\omega_{l}+\hbar^2q^2/(2mS(q))}~.
\label {4r} 
\end {equation} 

The latter is a perturbative method of treating the coupled 
e-ph system and it goes beyond the lowest order processes by using the R.P.A. 
for the e-e potential which contains the coulomb repulsion and the e-e 
interaction mediated by a single phonon exchange: 
\begin {equation} 
V(q,\omega)=v^{\infty}_{q}+M^{2}_{q} D^{0}(q,\omega) 
\label {5r} 
\end {equation} 
where $D^{0}(q,\omega)$ is the free LO phonon propagator: 
\begin {equation} 
D^{0}(q,\omega)=\frac{1} {\omega-\omega_l+i\eta}-
\frac{1} {\omega+\omega_l+i\eta}~.
\label {6r} 
\end {equation}
It is worth noting that the use of the R.P.A. describes the screening of 
both Coulomb e-e and Fr\"ohlich e-ph interactions, which are treated on the 
same footing.
In the following we will refer to these two models making use of the 
names variational approach and perturbative approach, respectively. 

The model, proposed in this paper, is obtained in the following way.
 
First we consider the Hamiltonian of two electrons interacting with the 
longitudinal optical phonons via the Fr\"ohlich coupling and repelling 
each other through the Coulomb force: 
\[ 
H=\frac{P^{2}}{2M}+\frac{p^{2}}{2\mu}+\frac{e^{2}}{\epsilon_{\infty}r}+ 
\sum_{\vec{q}}\hbar\omega_{l}a^{\dagger}_{\vec{q}}a_{\vec{q}} 
\] 
\begin {equation} 
+\sum_{\vec{k}}\left[M_{k}\rho_{\vec{k}}(\vec{r})e^{i\vec{k}\cdot\vec{R}}
a_{\vec{k}}+\text{h.c.}\right] 
\label {7r} 
\end {equation} 
where $\rho_{\vec{k}}(\vec{r})=(e^{\frac{i}{2}\vec{k}\cdot\vec{r}}+
\text{h.c.})$, $\vec{R}$, $\vec{P}$, $\vec{r}$, $\vec{p}$ are the position 
and momentum of the centre of mass of the pair and of the relative 
particle, $M$ and $\mu$ are the total and reduced masses respectively.
Then we obtain an effective potential for the two electrons of eq.(7) 
eliminating, as we will review briefly, the phonon degrees of freedom. 
Finally the many body effects are studied considering a system of electrons 
interacting with this effective potential.   
 
The Hamiltonian (7) commutes with 
\begin {equation} 
\vec{P}_{t}=\vec{P}+\sum_{\vec{k}}\hbar\vec{k}a^{\dagger}_{\vec{k}}a_{
\vec{k}} 
\label {8r} 
\end {equation} 
which is the total momentum of the system. 
The conservation law of the total momentum is taken into account through 
the unitary transformation: 
\begin {equation} 
U=\exp{[i(\vec{Q}-\sum_{\vec{k}}\vec{k}a^{\dagger}_{\vec{k}}a_{\vec{k}})
\cdot\vec{R}]} 
\label {9r} 
\end {equation} 
where $\hbar\vec{Q}$ is the eigenvalue of $\vec{P}_{t}$. 
Following LLP\cite{lee} we choose the variational trial 
ground state\cite{cataudella,bassani,altri}
for the transformed Hamiltonian $H_1=U^{-1}HU$: 
\begin {equation} 
|\psi>=U_{1}(\vec{r}) |0>\varphi(\vec{r}) 
\label {10r} 
\end {equation} 
where $|0>$ is the vacuum of $a_{\vec{k}}$ and the operator $U_{1}$ 
is given by 
\begin {equation} 
U_{1}(\vec{r})=\exp{[\sum_{\vec{k}}(f_{\vec{k}}(\vec{r})a_{\vec{k}}-
f^*_{\vec{k}}(\vec{r})a^{\dagger}_{\vec{k}})]}~. 
\label {11r} 
\end {equation} 
The envelope function $\varphi(\vec{r})$ is chosen to be an 
hydrogenic-like radial wave function: 
\begin {equation} 
\varphi(\vec{r})=\left[\frac{\left(2\gamma\right)^{2\beta+3}}{\Gamma(2\beta+3)}
\right]^{1/2}r^{\beta}e^{-\gamma r}
\label {12r} 
\end {equation} 
where $\Gamma(x)$ is the Gamma function. 
The phonon distribution functions $f_{\vec{k}}$ are determined in a 
self-consistent way from a functional variational procedure. In particular, 
the Eulero-Lagrange equations for the functions $f_{\vec{k}}$: 
\begin {equation} 
\frac{\hbar^2 k^2}{2 M} 
f_{\vec{k}}-\frac{\hbar^2}{2 \mu} \nabla^{2}{f_{\vec{k}}}+
\hbar\omega_l f_{\vec{k}}-\frac{\hbar^2}{2\mu} \nabla{f_{\vec{k}}} \cdot 
\frac{\nabla{\left|\varphi\right|^2}}{\left|\varphi\right|^2}=\rho_{\vec{k}} 
M_{\vec{k}}
\label {50r} 
\end {equation} 
can be solved exactly for fixed values of the 
parameters $\beta$ and $\gamma$: the solutions are expressed in terms of the 
regular and irregular confluent hypergeometric functions
\cite{cataudella,bassani,altri}. 
Then the parameters $\beta$ and $\gamma$ 
of the pair envelope function are fixed, in a variational way, 
by imposing the total energy, $\epsilon_t$, to be at a minimum.
Since $\epsilon_t$ reduces to that of two free polarons in the 
LLP approximation\cite{lee} when the average distance between the 
particles is much larger than the polaron radius ($\gamma=\beta=0$), 
the bipolaron binding energy 
is calculated subtracting from the total energy that of two LLP free polarons. 
In particular it has been shown that the bipolaron state exists only if the 
e-ph coupling constant $\alpha$ is greater than a critical value $\alpha_c=6$ 
and when $\eta=\epsilon_{0}/\epsilon_{\infty}$ is smaller than a 
critical value $\eta_c=0.01$\cite{cataudella,bassani}. 

It is worth to note that in the intermediate 
coupling regime it is not possible to neglect the $f_{\vec{k}}$ dependence on 
$r$. In this case the phonons follow instantaneously the relative motion of the 
two electrons and, then, the wave function (10) contains, on average, the 
retardation effects of the e-e interaction mediated by the longitudinal 
optical phonons. 

Within this variational approach it is 
possible to obtain an effective e-e potential due to the exchange of virtual 
phonons. 
In fact, the minimization procedure gives rise to two coupled differential 
equations for $f_{\vec{k}}$ and $\varphi$ which are solved self-consistently. 
With the choice (12) for the envelope function it is possible, as mentioned, 
to solve exactly the differential equation for $f_{\vec{k}}$ as a function 
of the variational parameters $\beta$ and $\gamma$. 
In the centre of mass frame the total energy minimization with respect to 
$\varphi(\vec{r})$, $\frac{\delta <\psi|H_1|\psi>}{\delta\varphi}$, 
gives a Schrodinger-like equation for the two electrons 
from which it is possible to define an effective potential depending by 
$\beta$ and $\gamma$ known the functions $f_{\vec{k}}$\cite{altri}: 
\begin {equation} 
v(r)=\frac{e^2}{\epsilon_{\infty} r}+
\frac{\hbar^2}{2\mu}\sum_{\vec{k}}\left|\nabla{ f_{\vec{k}} }\right|^2+
\sum_{\vec{k}}\left[\left(\frac{\hbar^2 \left|k\right|^2}{2 M}
+\hbar\omega_0\right)\left|f_{\vec{k}}\right|^2\right]-
\sum_{\vec{k}}\left[\rho_{\vec{k}}M_k f_{\vec{k}}+c.c.\right]~.
\label {61r} 
\end {equation} 

Typical effective potentials $v(r)$ are shown in Fig.1. 
They contain a short range attractive term and a long range 
repulsive term screened at large distances by the static dielectric 
constant $\epsilon_{0}$ and, in the opposite limit, by the background high 
frequency dielectric constant $\epsilon_{\infty}$. 
It is worth to note that the proposed approach can be, in principle, improved 
if one chooses better and better estimate for the effective potential. 

For the polar materials that are characterized 
by a value of $\alpha$ smaller than $\alpha_c$, there is not bipolaron 
formation and the variational approach gives $\gamma=\beta=0$. 
Consequently $v(r)$ 
coincides with the asymptotic effective e-e potential, i.e., with the 
effective potential of two electrons whose relative average distance is much 
larger than the polaron radius.  
On the other hand, for $\alpha > \alpha_c$ the effective potential supports the 
bipolaron formation. However, it is worth to note that bipolaron formation 
in the limit of vanishing density does not imply the existence of bipolarons 
at finite densities.

The proposed procedure allows us to eliminate the phonon degrees 
of freedom from the system, simplifying the treatment of many electron effects, 
and to investigate larger values of the electron-phonon coupling constant 
with respect to the perturbative approach proposed by Mahan. 
This happens since the LLP transformation\cite{lee} gives rise to phonon 
corrections to the bare e-ph vertex\cite{corrections}, corrections which can 
be neglected according to the Migdal theorem\cite{migdal} only when 
the Fermi energy is much larger than $\hbar\omega_l$ (normal metals). 
 
As mentioned before our approach contains two main approximations: a) 
the e-e and e-ph interactions are not treated on the same footing; 
b) the variational effective potential which contains the e-ph effects is not 
$\omega$-dependent as expected, for instance, in the usual Eliashberg theory 
of superconductivity\cite{eliasberg}. 
However, these two approximations are not significant limitations for the 
physical situation which are aimed to describe. For low values of the 
electron density (low doping) 
and for both weak and intermediate e-ph interactions ($\alpha<7$) 
it is clear that the 
approach is correct since the e-e effects are less important compared to the 
e-ph effects and the physics is controlled by the formation of polarons and, 
if it is possible, bipolarons well described by the LLP variational approach 
that we adopted. 
A mention deserves the strong e-ph coupling regime ($\alpha>7$)\cite{pekar}. 
Recently it has been shown\cite{quemerais}, 
by using the path integral technique, that 
the Wigner crystallization of polarons is favoured with respect 
to the metallic state up to critical electron density, $n_c$, of the order of 
$n_c\sim 10^{19}$ cm$^{-3}$. For $n>n_c$ the system of dielectrics polarons 
cannot form a liquid state at zero temperature and the transition to the 
metallic phase is driven by the polaron dissociation. In this work the regime 
characterised by very strong e-ph interactions will not be analyzed. 

The regime of larger doping but still such that 
($\omega^{0}_{p}<\omega_l$, $\omega^{0}_{p}$ being the plasma frequency 
screened by $\epsilon_{0}$)
and weak e-ph interaction also does not present 
any problem; in fact, as we will discuss later, our approach is equivalent to the 
available theories that do not suffer from the restrictions due to the 
approximations a) and b).
 
The validity of our approach is less evident for intermediate 
e-ph interaction and $\omega^0_p\sim\omega_l$. 
In fact, it is well known that the electron 
corrections to the bare e-ph vertex tend to suppress  the 
effective phonon-mediated e-e interaction when 
$v_{\scriptscriptstyle{F}}q/\omega \gg 1$, 
$q$ and $\omega$ being the transferred momentum and 
energy. As a consequence, our approach should overestimate e-ph 
effects unless $\omega$ is very large. 
However, when the system is close to a charge-density waves instability, 
it has been shown that the vertex 
corrections due to the e-e repulsion are ineffective also in the limit 
$v_{\scriptscriptstyle{F}}q/\omega\gg 1 $, while the phonon vertex 
corrections are particularly relevant\cite{dicastro}. 
In other words, it is a reasonable 
approximation to treat first the phonon corrections to the e-ph vertex and then 
to consider the effects due to the presence of many electrons. 
This recovers the validity of our approach.

Finally we note that at very large values of the carrier density, typical of 
the ordinary metals, 
the proposed approach is 
not able to describe the physics correctly since the retardation effects of the 
effective phonon-mediated e-e are relevant, electron screening very effective 
and the phonon vertex 
corrections are negligible according to the Migdal theorem\cite{migdal}.  

The model proposed is studied within the R.P.A.\cite{lindhard} and Hubbard
approximation \cite{hubbard} at $T=0$. Within these approximations the 
effective interaction between the 
electrons takes the form: 
\begin {equation} 
V_{eff}(q,\omega)=\frac{v(q)}{1-V^*(q,\omega)/[1+f(q)V^*(q,\omega)]}
=\frac{v(q)}{\epsilon(q,\omega)} 
\label {13r} 
\end {equation} 
where $V^*(q,\omega)=v(q)\Pi^{0}(q,\omega)$ and
$\Pi^{0}(q,\omega)$ is the lowest order proper polarization 
propagator. The function $f(q)$ takes the value $0$ in the R.P.A.
\cite{lindhard} and 
\begin {equation} 
f(q)=\frac{1}{2}\frac{v\left(\sqrt{q^{2}+q^{2}_{F}}\right)}{v(q)} 
\label {14r} 
\end {equation} 
in the Hubbard approximation \cite{hubbard}. 
The R.P.A. excluding exchange diagrams overestimates the contribution 
of the short range terms by around a factor of two. The Hubbard 
approximation takes into account approximatively these terms making an estimate 
of the contribution of all proper polarization insertions with repeated 
horizontal interaction lines across the fermion loop. 
In the limit of large electron densities the Hubbard approximation 
coincides with the R.P.A., the dominant contribution to the correlation 
energy coming from low momentum transfers. 

The calculation of the self-energy to the lowest order in the dynamically 
screened interaction $V_{eff}(q,\omega)$ gives: 
\begin {equation} 
\Sigma(k,\omega)=\frac{i}{\hbar\left(2\pi\right)^{4}}
\int_{0}^{\infty} d\omega_{1}
\int d\vec{q} ~~G^{0}(\vec{k}-\vec{q},\omega-\omega_{1})\frac{v(q)}
{\epsilon(q,\omega_1)} 
\label {16r} 
\end {equation} 
where $G^{0}(q,\omega)$ is the propagator for non interacting electrons 
\cite{fetter} 
\begin {equation} 
G^{0}(q,\omega)=\frac{\Theta(q-q_{\scriptscriptstyle{F}})}
{\omega-E^0_q/\hbar-\epsilon^*_{0}
+i\eta}+\frac{\Theta(q_{\scriptscriptstyle{F}}-q)}
{\omega-E^0_q/\hbar-\epsilon^*_{0}-i\eta}~. 
\label {17r} 
\end {equation} 
In the following we will focus our attention on the properties of the 
normal 
state: therefore the energy shift $\epsilon^*_{0}$ is 
chosen 
so that the chemical potential of the non interacting system $\mu= 
\hbar^{2}k^{2}_{F}/2m+\epsilon^*_{0}$ coincides with that obtained by 
the equation\cite{hedin,lundqvist}: 
\begin {equation} 
\mu=\frac{\hbar^{2}k^{2}_{F}}{2m}+\Sigma(k_{F},\mu-\epsilon^*_{0}) 
\label{18r} 
\end {equation} 
Eq.(\ref{18r}) combined with the above expression for $\mu$ gives: 
\begin {equation} 
\epsilon^*_{0}=\Sigma(k_{F},\frac{\hbar^{2}k^{2}_{F}}{2m})~. 
\label {19r} 
\end {equation} 
Finally to evaluate the self-energy we follow the treatment proposed 
by Quinn and Ferrel\cite{quinn}, which retains all contributions coming from 
the continuum of the electron-hole pair states.
 
\section {Weak coupling} 

The aim of this section is to compare a number of properties, obtained within 
the model 
introduced before, with the results given by the variational and perturbative 
approaches. 
In particular we will focus our attention on the total energy, the 
collective excitations and the single-particle self-energy: numerical 
results will be presented for $ZnO$, which is characterised by the 
following parameters\cite{kartheuser}: $m=.24m_{e}$, $\epsilon_{\infty}=4$,
$\epsilon_{0}=8.15$, $\hbar\omega_{l}=73.27\, meV$, $\alpha=.849$. 

\subsection{The total ground state energy}

The total ground state energy per particle takes the following 
form: 
\begin {equation} 
E_{T}=T_{0}+E_{ex}+E_{corr} 
\label {20r} 
\end {equation} 
where 
\begin{eqnarray}
&T_{0}=\frac{3}{5} E_F,  ~~  
E_{ex}=-\frac{Vk^{3}_{F}}{12N\pi^{4}}\int_{0}^{2k_{F}}dq\, q^{2}v(q)
\left[1-1.5\frac{q}{2k_{F}}+.5\left(\frac{q}{2k_{F}}\right)^{3}\right]
\nonumber\\
&\nonumber\\
&E_{corr}=\frac{3e^2}{4a_B\pi\alpha^{2}_{1}r^{2}_{s}}\int_{0}^{\infty}dq\, 
\frac{q^{2}}{2k^3_F}\int_{0}^{\infty}\frac{d\omega}{2E_F}\left[\frac{1}{g(q)}
\arctan{\left(\frac{v(q)\text{Im}\Pi^{0}(q,\omega)g(q)}
{1-g(q)v(q)\text{Re}\Pi^{0}(q,\omega)}\right)}
-v(q)\text{Im}\Pi^{0}(q,\omega)\right] \nonumber\\ 
&\nonumber\\
&\alpha_{1}=\left(\frac{4}{9\pi}\right)^{1/3}, ~~ g(q)=1-f(q)~. 
\end{eqnarray}

In Eq.(\ref{20r}) $T_{0}$ and $E_{ex}$ are the kinetic and potential 
energy, respectively, in H.F. approximation; $E_{corr}$ is the 
correlation energy in the Hubbard approximation and $r_{s}$ is the radius, 
in the Bohr units, of a sphere whose volume is equal to the volume per 
particle. In particular in the limit $\epsilon_{0}=\epsilon_{\infty}$ and 
$m=m_{e}$ eq. (\ref{20r}) provides the total ground state energy per 
particle of a Coulomb gas. 
In Fig.2a  the contribution of the e-ph interaction to the 
ground state energy in the variational\cite{costa} and our approaches are 
plotted as a function of the electron density. In the variational approach 
this contribution is given by $E_p$ in Eq. (\ref{3r}). Since in our model it 
is not possible to express the total ground state energy as the 
sum of two independent terms, the e-ph contribution to $E_T$  is 
obtained subtracting from $E_T$ the energy per particle of a Coulomb gas 
screened by the background high frequency dielectric constant 
$\epsilon_{\infty}$. 
In both models, increasing 
the density, $|E_{p}|$ decreases, the screening of the Coulomb gas on the 
polaron self-energy becoming bigger and bigger. However, in our 
approach the effect of the electron screening is slightly less effective for 
low values of the doping. Instead, for very large electron densities, our model 
underestimates the effects of the e-ph interaction. In Fig.2b the same 
quantity is plotted as a function 
of $\alpha$ for a fixed value of the electron density, $n=10^{18}$ cm$^{-3}$. 
It is evident that the two approaches provide the same results when the e-ph 
coupling constant is very little ($\alpha<1$).

\subsection                 { Collective excitation spectrum}
 
The collective excitation frequencies of the system are determined by the 
poles of the retarded density correlation function, which occur at the 
solutions $\Omega_{q}-i\gamma_{q}$ of the equation: 
\begin {equation} 
\epsilon^{R}(q,\Omega_{q}-i\gamma_{q})=0~. 
\label {21r} 
\end {equation} 

In the perturbative approach the dielectric function is chosen to be the 
sum of three contributions\cite{mahan}: 
\begin {equation} 
\epsilon(q,\omega)=\epsilon_{\infty}+\frac{\epsilon_{0}-\epsilon_{\infty}}
{1-\omega^{2}/\omega^{2}_{T}}-v^{\infty}_q\Pi^{0}(q,\omega) 
\label {22r} 
\end {equation} 
where $\omega_{T}$ is the transverse optical phonon frequency. 
This expression, which is exact within the R.P.A., takes 
into account the screening effects due to the electron gas, the optical 
phonons and the high energy electronic excitations. 
Eq.(\ref{22r}) is a quadratic equation for $\omega^{2}$ in the limit 
$q\rightarrow 0$: therefore there are always two roots $\omega_{1}^{2}(q)$ 
and $\omega_{2}^{2}(q)$ that are a mixing of the phonon and plasmon 
excitations. There are two opposite limits in which the many-polaron effects 
can be easily treated: they occur when the plasma frequency 
$\omega^{\infty}_{p}$ (the plasma frequency screened by 
$\epsilon_{\infty}$)
is much smaller or much larger than $\omega_{l}$. 

In the former case $\omega_1$ tends to $\omega_l$, the 
electron gas being not able to oscillate as fast as the phonons. On the other 
hand, since the phonons can follow the motion of the electrons 
$\omega_2(q=0)$ tends to the plasmon frequency screened by $\epsilon_0$. 
The situation is different when $\omega_p^{\infty}$ is much larger $\omega_l$. 
In fact  the electrons see the ionic motion as static so that 
the static electron screening can be used. In this case the frequencies 
$\omega_1(q)$ and $\omega_2(q)$ have a different behaviour depending on the 
$q$ value. In fact the electron gas has a 
characteristic screening length $q^{-1}_{\scriptscriptstyle{TF}}$, 
$q_{\scriptscriptstyle{TF}}$ being the Thomas-Fermi wave vector, and the 
electron-electron interaction declines rapidly 
at distances larger than the screening length, thus for 
$q>q_{\scriptscriptstyle{TF}}$ the 
screening of the electron gas becomes less effective. For 
$q \ll q_{\scriptscriptstyle{TF}}$ 
the mode $\omega_{2}^2(q)$ tends to $\omega^2_{T}$ while at large values of 
$q$ it goes to $\omega^2_{l}$ since the electron gas is not any longer able 
to screen the electron-phonon interaction. 
The phonons cannot follow the plasma oscillations of the electron gas: then 
the plasma frequency is screened by the dielectric constant 
$\epsilon_{\infty}$. 

In our approach the description of the collective excitation spectrum is less 
rich. In fact since 
the retardation effects of the LO phonon-mediated e-e interaction have been 
neglected, there is always only one solution to Eq.(\ref{21r}). 
The collective excitation energy goes from $\hbar\omega_{p}/
\sqrt{\epsilon_{0}}$ for 
$q\rightarrow 0$ to the roots of an electron gas screened by the 
background high frequency dielectric constant $\epsilon_{\infty}$ 
for large values of $q$. 
Particularly at low values of the electron density this mode coincides 
with the plasmon like mode in the perturbative approach. 
We also note that, in our approach, it is not possible to divide the 
dielectric function into different contributions due to the e-e interaction 
and e-ph interaction.

\subsection             {Imaginary part of the self-energy }

In Fig.3 we show the results of the imaginary part of the electron 
self-energy $\text{Im}\Sigma(k,\omega)$ in our model, as a function of 
$\omega$ in the Hubbard approximation. $\text{Im}\Sigma(k,\omega)$ is 
zero for all $k$ at $\hbar\omega=E_{F}$: a change of sign must 
occur at this point since the damping of electrons and holes is 
opposite in sign. The large peak contribution to $\text{Im}\Sigma
(k,\omega)$ is due to the excitation of the collective mode. 
Since the real and 
imaginary parts of $\Sigma(k,\omega)$ obey the Kramers-Kronigh dispersion 
relation, there is a corresponding structure in $\text{Re}\Sigma(k,\omega)$: 
it has a finite discontinuity at the same point in which $\text{Im}\Sigma
(k,\omega)$ shows a peak. The other contribution to $\text{Im}\Sigma
(k,\omega)$ arises from the region where $\epsilon_{2}(k,\omega)$ is 
not zero, $\epsilon_{2}(k,\omega)$ being the imaginary part of the 
dielectric function, and it is due to the creation of particle-hole 
pairs. 
It is also possible to evaluate the imaginary part of the self-energy 
in the perturbative approach in which the effective potential between the 
electrons is the sum of a) the screened coulomb interaction and b) the 
screened electron-phonon interaction. The b) contribution is expressed as the 
product of the screened matrix element $M^{2}_{q}/(\epsilon^{2}_{el}(q,\omega))
$ and the phonon Green's function $D(q,\omega)$: 
\begin {equation} 
D(q,\omega)=\frac{D^{0}(q,\omega)}{1-M^{2}_{q}D^{0}\Pi^{0}
/\epsilon_{el}(q,\omega)} 
\label {23r} 
\end {equation} 
where 
\begin {equation} 
\epsilon_{el}(q,\omega)=1-v^{\infty}_q\Pi^{0}(q,\omega)~. 
\label {24r} 
\end {equation} 
Mahan has calculated the one-phonon self-energy term from the electron-
phonon interaction replacing the phonon Green's function $D(q,\omega)$ 
by the unperturbated propagator $D^{0}(q,\omega)$ and making use of the 
static dielectric function of Thomas-Fermi. The result is that $\text
{Im}\Sigma(k,\omega)$ is zero for energies within $\omega_{l}$ around the Fermi 
energy. The onset of $\text{Im}\Sigma(k,\omega)$  for $\omega>\omega_
{l}$ corresponds to the fact that the phonon emission processes are not 
possible until the energy is equal or greater than the phonon frequency 
$\omega_{l}$. 
Moreover at $T=0$, if the system is in equilibrium, the contribution to 
the imaginary part of the self-energy from the electron-phonon interaction 
becomes again zero for $\hbar\omega<-\hbar\omega_{l}-E_{F}$. The onset of 
$\text{Im}\Sigma(k,\omega)$ at $\hbar\omega=-\hbar\omega_{l}-E_{F}$ 
corresponds to the process in which the electron with energy $-E_{F}$, 
jumps in the hole having energy $-\hbar\omega_{l}-E_{F}$. 
These results were first obtained by Engelsberg and Schrieffer\cite
{engelsberg} who considered a system of electrons coupled with the acoustical 
phonons. 
We have calculated the self-energy  within the perturbative approach making 
use of the R.P.A. approximation and the 
results are presented in Fig.4: besides the peak, due to the excitation of the 
coupled phonon-plasmon collective mode, the 
imaginary part of the electron self-energy shows a strong variation in 
the points $+\hbar\omega_{l}$, $-\hbar\omega_{l}$ and $-\hbar\omega_{l}-E_{F}$,
but the amplitude of the gap coincides with that obtained making use of the 
Thomas-Fermi approximation for the dielectric constant only in the limit 
$\omega^{\infty}_{p} \gg \omega_{l}$. 
These results show, as it is reasonable, that the static approximation 
provides a good treatment of the electronic effects only when the 
plasma frequency is much larger than the characteristic phonon energies. 
Similar results have been obtained by Das Sarma et al.\cite{das sarma} 
who have calculated the quasi-particle damping rate which is 
proportional to the quantity $\Gamma(k)=\text{Im}\Sigma(k,k^{2})$. They 
showed that the static screening is a good approximation at very high 
electron densities, whereas the dynamically screened decay rate lies 
intermediate between statically screened and unscreened results. 
In Fig.5 $\Gamma(k)$ is plotted as  function of $k$ for the $ZnO$ at 
$n=10^{17}$ cm$^{-3}$ ($\omega^{\infty}_{p}\ll\omega_{l}$), 
both in our model and in the perturbative approach making use of the R.P.A.. 
In particular in the region close to the Fermi surface 
the damping rate goes like $\left(E^{0}(k)-\mu\right)^{2}$ where $E^{0}
(k)$ is the quasi-particle energy: this is a simple consequence of the 
Pauli principle restrictions. The peaks in $\Gamma(k)$ are caused by 
processes corresponding to the excitations of the two renormalized modes 
$\omega_{1}(k)$ and $\omega_{2}(k)$. 
Moreover it is evident that at low electron densities $\omega^{\infty}_
{p}\ll\omega_{l}$ our model neglects, with respect to the perturbative 
approach, excitation processes involving high energy, whereas 
it provides a good treatment of the characteristic polaronic energies. 
We draw the same conclusions evaluating the polaronic corrections to the 
band edge which, in the Rayleigh-Schrodinger perturbation theory, are 
given by $E_{p}=\text{Re}\Sigma_1(k=0,\omega=0)$, $\Sigma_1(k,\omega)$ 
being the one-phonon self-energy term from the e-ph interaction. 
In Fig.6 we present the results 
for $E_{p}$ in the perturbative approach making use of the R.P.A., static 
R.P.A. and the Thomas-Fermi approximations. 
It is clear that the Thomas-Fermi approximation overestimates the 
screening effects at low densities, giving a smaller polaronic binding 
energy than the dynamically screened results. In the opposite limit, 
instead, the two approximations provide the same results.

\subsection    {The spectral weight function} 

The spectral weight function is the imaginary part of the electron 
propagator and therefore it can be expressed in term of the real and 
imaginary parts of the self energy $\Sigma_{ret}(k,\omega)$\cite{nota1} 
\begin {equation} 
A(k,\omega)=-\frac{2\text{Im}\Sigma_{ret}(k,\omega)}{\left[\omega-E^0_k/\hbar
-\text{Re}\Sigma_{ret}(k,\omega)\right]^{2}+\left[\text{Im}\Sigma_{ret}
(k,\omega)\right]^{2}}~. 
\label {25r} 
\end {equation}
 
The Fig.7a, 7c show $A(k,\omega)$ in the Hubbard approximation for two 
different values of $\eta$. The energy $E(k)$ 
and the damping rate $\Gamma(k)$ of a quasi-electron ($k>k_F$) and a quasi-hole 
($k<k_{F}$) are the poles of the analytic continuation of $G(k,\omega)$ 
into the lower right and upper left half plane for $\omega$ : 
\begin {equation} 
E(k)=\frac{\hbar^{2}k^{2}}{2m}+\text{Re}\Sigma\left(k,\frac{E(k)+i\Gamma(k)}
{\hbar}\right) 
\label {26r} 
\end {equation} 
and 
\begin {equation} 
\Gamma(k)=\text{Im}\Sigma\left(k,\frac{E(k)+i\Gamma(k)}{\hbar}\right)~. 
\label {27r} 
\end {equation} 
In particular for long lived single particle excitations the energy $E(k)$ 
and the damping rate $\Gamma(k)$ are given by: 
\begin {equation} 
E(k)=\frac{\hbar^{2}k^{2}}{2m}+\text{Re}\Sigma\left(k,\frac{E(k)}{\hbar}
\right)
\label{28r} 
\end {equation} 
and 
\begin {equation} 
\Gamma(k)=\frac{\text{Im}\Sigma(k,\frac{E(k)}{\hbar})}
{\left.\left(1-\frac
{\partial\text{Re}\Sigma(k,\omega)}{\partial \omega}\right)\right|_{\frac{
E(k)}{\hbar}}}~. 
\label {29r} 
\end {equation} 
The numerical solution of Eq.(\ref{28r}) shows the existence of three 
roots. The one at highest energy is the regular quasi-particle solution 
(screened polaron), which 
is very little shifted from the Hartree-Fock value. 
It describes electrons surrounded by a dynamic polarization cloud due to the 
effective e-e interaction: the self-energy at $q=0$ of this quasi-particle 
increases with the particle density. 
Another solution appears at the finite discontinuity of $\text{Re}\Sigma
(k,\omega)$ and it doesn't give contribution to the spectral weight 
function having the imaginary part of the electron self-energy a peak at 
the same energy. To understand the physical meaning of the other 
solution it is convenient to separate the contribution of the 
exchange term from the self-energy: 
$\Sigma(k,\omega)=\Sigma_{exc}(k,\omega)+\Sigma^*(k,\omega)$. 
The contribution given by $\Sigma^*(k,\omega)$ can be viewed as that of a 
particle coupled to the density fluctuations propagator $S(k,\omega)$, 
whose spectral representation is: 
\begin {equation} 
S(k,\omega)=-\frac{1}{\pi v(k)}\int_{0}^{\infty}d\omega_{1} \text{Im}
\frac{1}{\epsilon(k,\omega_{1})} \frac{2\omega_{1}}{\omega^{2}-
\left(\omega_{1}-i\delta\right)^{2}}~. 
\label {30r} 
\end {equation} 

On the other hand, the single-particle self-energy  
of a fermion coupled to a boson field with the Bose propagator: 
\begin {equation} 
D(k,\omega)=\frac{2\omega(k)}{\omega^{2}-\omega^{2}(k)+i\delta} 
\label {31r} 
\end {equation} 
and the effective coupling: 
\begin {equation} 
g^{2}(k)=\frac{v(k)} 
{\left|\frac{\partial\epsilon(k,\omega)}{\partial
\omega}\right|_{\omega(k)}} 
\label {32r} 
\end {equation} 
takes exactly the form $\Sigma^*(k,\omega)$ if the further approximation 
of single plasma pole 
for the density fluctuations $S(k,\omega)$ is used \cite{lundqvist}. 
Therefore $\Sigma^*(k,\omega)$ can be approximately described as the 
contribution of a particle interacting with a boson field: the plasmon-phonon. 
The quasi-particle formed by an electron surrounded by a cloud of virtual 
excitations (plasmon-phonon) is usually called plasmapolaron \cite{iadonisi}.

The quantity $E-\epsilon^*_{0r}$, where $E$ is the $q=0$ self-energy of 
the plasmapolaron, i.e. the solution of the equation $E=\text{Re}\Sigma
(0,E)$, and $\epsilon^*_{0r}$ is the shift in the chemical potential for a
Coulomb gas screened by $\epsilon_{\infty}$,  
is plotted as a function of the electron density in Fig.8. 
It is in good agreement with that calculated starting by the Hamiltonian 
of an electron interacting with the phonon and plasmon boson fields and 
using a variational approach based on the LLP 
transformation\cite{iadonisi}.
The differences between the two plots are due to the presence, in the 
self-energy, of the continuum of the electron-hole pair states, exchange and 
correlation terms. These contributions are not considered in the variational 
approach proposed in ref.[37]. 

In Fig.7b, 7d the spectral weight function in the perturbative approach is 
plotted for the same values of $\eta$, $n$ and $k$. 
It is well known that, in the perturbative approach, $A(k,\omega)$ 
for a single electron coupled to the longitudinal optical phonons shows,  
at $T=0$ and for $k=0$, a delta-function peak and an incoherent contribution, 
with very small spectral weight, which starts at $\omega=\omega_l$\cite{mahan}. 
The effect of the e-e interaction is to split the single delta-function peak 
in two lorentian functions that describe two quasi-particle: the polaron 
and the plasmapolaron. The 
comparison between Fig.7a and Fig.7b points out, again, that our approach 
takes well into account the characteristic polaronic energies and neglects the 
excitation processes involving high energies ($\omega>\omega_l$). This is a
reasonable approximation when $\omega^0_p<\omega_l$. 

It is worth to note that, increasing the charge carrier density, the polaron, 
an electron dressed by a cloud of virtual phonons, evolves towards a new 
quasiparticle, an electron weakly renormalized by the scattering with 
the phonons, i.e., a quasiparticle with reduced energy and mass with respect 
to the values 
$E_p(q=0)=-\alpha\hbar\omega_l$ and $m^*=m/\left(1-\alpha/6\right)$. 
Instead the plasmapolaron evolves towards an electron dressed by a 
cloud of virtual plasmons. 
   
\section  {Intermediate coupling $(\alpha \sim\alpha_{\scriptscriptstyle{c}})$} 

\subsection                 { Collective excitation spectrum}

It is well known that the retarded dielectric function is analytic in the 
upper half of the complex plane provided\cite{pines} 
$\epsilon^{R}(k,0) \geq 0$. 
There are examples of physical systems for which the linear response 
function violates the causal requirement: in this case the linear retarded 
dielectric function can no longer describe the behaviour of the system 
correctly. It has been shown that the temperature dependent correlation 
function in the R.P.A. 
for an interacting many-particle system, $\epsilon^{R}(k,\omega)$, when 
the interaction is sufficiently attractive, has forbidden zeros on the 
imaginary axis of the complex frequency plane\cite{mermin}. These poles 
of the linear response function appear even in the classical limit where 
they correspond to the transition of the system from a gaseous to a liquid 
state. A similar situation is found in the the case of a superconductor
\cite{kadanoff}. 
While in the perturbative approach the causal requirement is never 
violated, even if the interaction electron-phonon is extremely strong, 
in our model, if the value of the coupling constant $\alpha$ is 
sufficiently large, the linear response function possesses a pair of 
imaginary poles. 
It is interesting to note that the same type of instability has been 
suggested by Di Castro et al. for a large class of systems of interacting 
electrons and phonons as due to very ineffective electron corrections to the 
e-ph vertex\cite{dicastro}. 

The results which will be shown here and in the following subsections 
have been obtained 
choosing the model parameters in such a way the system is just above the 
appearance of the instability. 
The parameters chosen are: 
$m=m_e$, $\eta=.01$, $R_{p}=10$ \AA, $\alpha\hbar\omega=.016~Ry$, 
($\alpha\sim\alpha_c$).

In Fig.9 we present the numerical results for the collective excitation 
spectrum in 
the Hubbard approximation. The collective mode energy softens for a 
critical wave vector q indicating strong correlation between the 
electrons. If the attractive potential is sufficiently strong the 
collective energy softens completely. This softening, present in a finite 
range of densities $ 10^{19}-10^{21}$ cm$^{-3}$, means that the system 
becomes unstable with respect to the formation of the charge density waves. 
Similar results are also obtained in the Hubbard model with on site 
attraction\cite{micnas} 
and using the slave-boson approach for the infinite U three band Hubbard 
model\cite{dicastro}. 

\subsection                 {The spectral weight function}
 
The Fig.10 shows $A(k,\omega)$ for $k=0.01k_{F}$ and for different values 
of the electron density in the Hubbard approximation. Beside the peaks 
already present in the weak coupling regime the Eq.(\ref{28r}) exhibits 
a new solution with energy between the plasmapolaron and quasi-
particle branches. It appears only if the strength of the attractive potential 
is sufficiently large and the damping rate of this excitation increases 
very quickly as a function of $k$. Then it exists as a well-defined 
excitation of the system only for moments well inside the Fermi surface. 
As mentioned before this excitation is related to the attractive part of the 
potential and it is present only close to the instability signaled by the 
complete softening of the collective mode energy. 
This interpretation is also confirmed by the behaviour of 
the renormalization coefficient of the one-electron Green's function: 
\begin {equation} 
Z(k)=\frac{1}{1-\frac{\partial\text{Re}\Sigma(k,\omega)}{\partial\omega}}~. 
\label {33r} 
\end {equation} 
At $k=k_{F}$, $Z(k_{F})$ is the magnitude of the discontinuity 
at $k=k_{F}$ in the momentum distribution $n(k)$\cite{luttinger}. 
In Fig.11 $Z(k_{F})$ is plotted as a function of the electron density and it 
is compared with the renormalization coefficient of a Coulomb gas screened 
by the background high frequency dielectric constant 
$\epsilon_{\infty}$\cite{rice}. 
In the range of densities in which the new elementary excitation appears, 
$Z(k_{F})$ has a minimum; increasing the electron density it tends 
to the value of a Coulomb gas screened by $\epsilon_{\infty}$. 
These results indicate that the charge density wave instability is a source of 
a strong quasi-particle scattering 
and suggest that the new peak in the spectral weight function 
could be a metastable double occupied state (bipolaron). 

\subsection                   { The dispersion curves} 

We end this section showing the dispersion curves of the three excitations 
$E_i(k)$ in the region of $k$ in which the decay is due only to the 
excitations of particle-hole pairs (Fig.12). They have been obtained following 
the poles of the analytic continuation of $G(k,\omega)$. 
In particular, for $k$ less than $k_{F}$ and energies less than $\mu$, these 
poles describe the 
state obtained by creating a hole in the interacting ground state. 
The curves $E_i(k)$ are drown only for k values which do not involve the 
creation of collective excitations. In fact the decay for 
the quasi-particle band due to the creation of collective excitations starts 
for a value of the electron momentum which satisfies the conservation of the 
energy: 
\begin {equation} 
\frac{\hbar^{2}k^{2}}{2m}=\frac{\hbar^{2}k^{2}_{F}}{2m}+\hbar\omega(k-k_
{F}) 
\label {34r} 
\end {equation} 
$\omega(k)$ being the collective excitation spectrum of the system, while 
the critical momentum which characterizes the decay of a plasmapolaron is given 
by the condition: 
\begin {equation} 
E(k)-\mu<-\hbar\omega(k+k_{F})~. 
\label {35r} 
\end {equation} 
Both the quasi-particle and plasmapolaron have approximatively 
quadratic dispersion laws.    

\section {Conclusions} 
A model for an interacting gas of large polarons has been introduced. In the 
model the electrons interact each other with a non retarded effective 
potential obtained within a variational approach\cite{cataudella} valid for 
large polarons in the intermediate e-ph coupling. The effective potential 
is able to recover the large polaron properties in the limit of zero density. 
Then the many-body problem has been studied within the R.P.A. and Hubbard 
approximations.  
The ground state energy, the single-particle self-energy and 
the collective excitation spectrum for a system of interacting large polarons 
have been calculated at densities typical of the metallic phase of doped polar 
semiconductors and doped perovskites. The numerical results have been compared 
with those provided by two traditional approaches in which 
the e-e and e-ph interactions are taken into account at the same level. 
It has been shown that the polaron self-energy is strongly 
reduced compared to that obtained in the zero density limit.
For intermediate values of the e-ph coupling constant the collective mode 
energy presents a softening in a 
finite range of densities at a finite value of $q$ signaling a CDW 
instability as suggested in recent works about the e-ph interaction in presence 
of strong correlations \cite{dicastro} and 
the single-particle spectrum shows a new
elementary excitation, a bipolaronic metastable state, with energy between the 
plasmapolaron and the quasi-particle branches. 
 
\section {acknowledgments} 
The authors are grateful to Prof. M. Marinaro for interesting discussions 
during the course of this work. GDF also thanks the European 
Economic Community for financial support. 
 
\section*{Figure captions} 
\begin {description} 
\item{Fig.1.} The effective self-consistent electron-electron potential. 
The energies are given in units of $\hbar\omega_l$ and are measured from 
$-2\alpha\hbar\omega_l$. The parameters of Fig.1a correspond to the 
characteristic values of $ZnO$. 
\item{Fig.2.} a) Electron-phonon contribution to the ground state energy per 
particle in the variational 
(dotted line) and in our model (solid line) as a function of the electron 
density. The energies are given in units of $\hbar\omega_l$; b) The same 
quantity is plotted for a fixed value of the electron density, 
$n=10^{18}$ cm$^{-3}$, 
as a function of the e-ph coupling constant. 
The parameters are the same of $ZnO$, except $\epsilon_{\infty}$, that varies 
from $1.4$ to $8.15$. 
\item{Fig.3.} The imaginary part of the self energy for $k=.5k_F$ as a 
function of the 
electron energy $\hbar\omega$ measured from $\epsilon^*_0$ at 
$n=10^{17}$ cm$^{-3}$. 
$\text{Im}\Sigma(k,\omega)$ is given in units of 
$4E_F$. 
\item{Fig.4.} The imaginary part of the self energy for $k=.5k_F$ as a 
function of the 
electron energy $\hbar\omega$ measured from $\epsilon^*_0$ at 
$n=10^{17}$ cm$^{-3}$ in the perturbative approach. In the region of the 
characteristic polaronic energies these results are very similar
to that obtained in our model reported in Fig.3. 
$\text{Im}\Sigma(k,\omega)$ is given in units of $4E_F$. 
\item{Fig.5.} The imaginary part of the quasi-particle self energy 
$\text{Im}\Sigma(k,k^2)$ as a function of the momentum $k$ of the 
electron at $n=10^{17}$ cm$^{-3}$ in our model (solid line) and in the 
perturbative approach (dotted line). The energies are given in units of 
$\hbar\omega_l$. 
\item {Fig.6.} The contribution of the electron-phonon interaction to the 
correction of the band edge in the perturbative approach making use of the 
R.P.A.(solid line),static R.P.A.(dashed line) and Thomas-Fermi (dotted line) 
approximations as a function of the electron density. 
The energies are given in units of $\hbar\omega_l$.
\item {Fig.7.} The one-particle spectral function 
in the range of the characteristic polaronic energies  
for $k=.5k_F$ and at  $n=10^{17}$ cm$^{-3}$, 
in our model and in the perturbative approach 
for two different values of $\eta$ (a, c and b, d  respectively).  
In both cases $\epsilon^*_0$ 
is of order of $\epsilon^*_0\sim-\alpha\hbar\omega_l$. 
The energies  are given in units of $4E_F$. The parameters of Fig.7a,7c 
correspond to the characteristic values of $ZnO$. The spectral weights 
relative to the different peaks are indicated. 
\item {Fig.8.}The plasmapolaron self-energy as a function of the electron 
density in our model (solid line) compared with that obtained in ref.[37]. 
The energies  are given in Ry and are measured from 
$-\alpha\hbar\omega_l$. 
\item {Fig.9.} The curves a) and c) represent the collective excitation
spectrum of the Coulomb gas screened by $\epsilon_0$ and $\epsilon_{\infty}$ 
respectively at different electron densities ($n=10^{19}$ cm$^{-3}$, 
$n=10^{20}$ cm$^{-3}$, $n=10^{21}$ cm$^{-3}$).
The 
particle-hole excitations lie between the dotted lines. The curve b) 
represents the energy of the collective mode as a function of the wave vector 
$q$ using the potential plotted in fig.1. 
The energies 
$\hbar\Omega(q)$ are given in units of $4E_F$. 
\item {Fig.10.} The one-particle spectral function for $k=.01k_F$ as a function 
of the electron energy at different charge carrier densities 
($n=10^{19}$ cm$^{-3}$, 
$n=10^{20}$ cm$^{-3}$). The energies  are given in units of $4E_F$.    
\item {Fig.11.} The renormalization coefficient of the one-electron Green's 
function at $k=k_F$ as a function of the density. The solid line represents 
$Z(k_F)$ for a Coulomb gas screened by the background high frequency 
dielectric constant $\epsilon_{\infty}$. 
\item {Fig.12.} The dispersion curves at different electron densities 
($n=10^{20}$ cm$^{-3}$, $n=10^{21}$ cm$^{-3}$) in the region of $k$ in which 
the decay is due only to the particle-hole pairs excitations. The energies 
are given in units of $4E_F$. 
\end {description}  
 
\end {document}
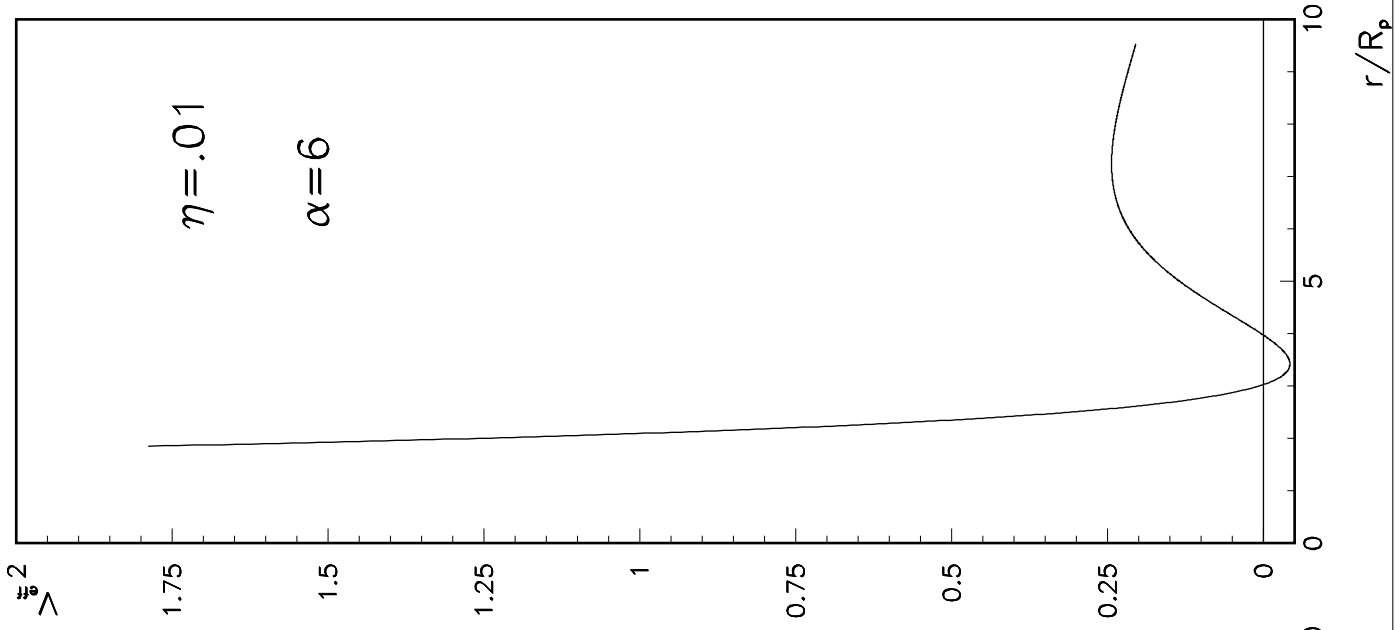
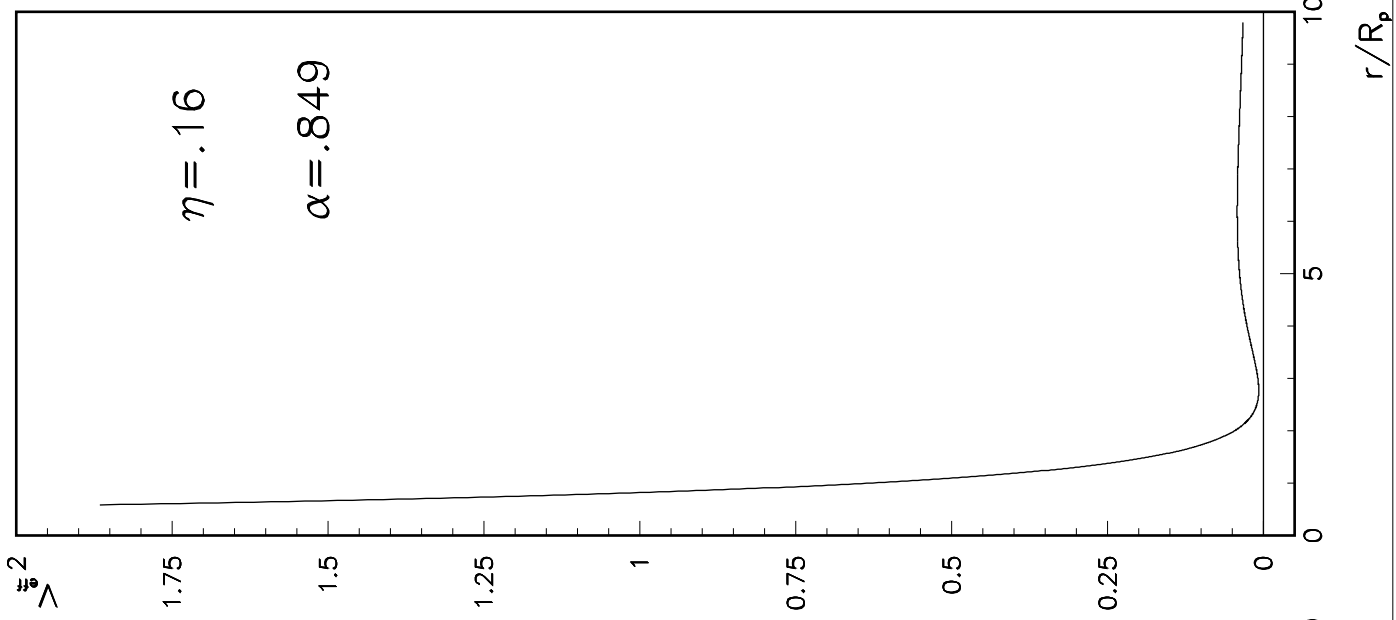
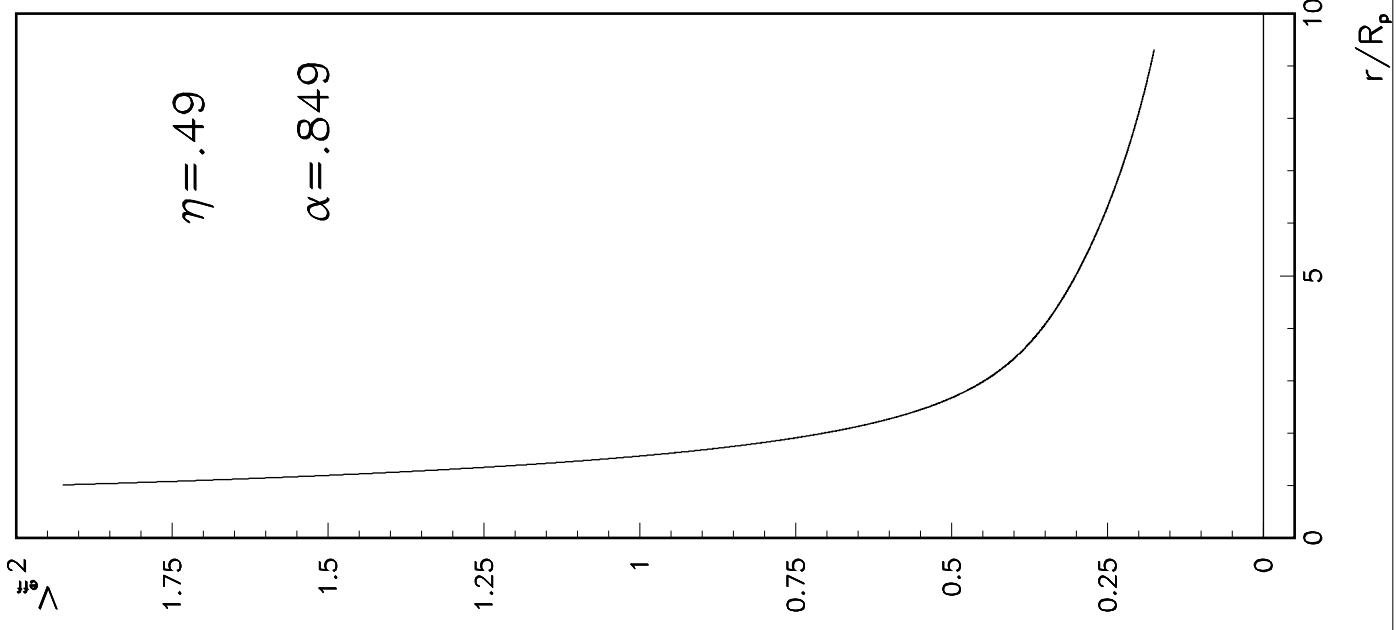
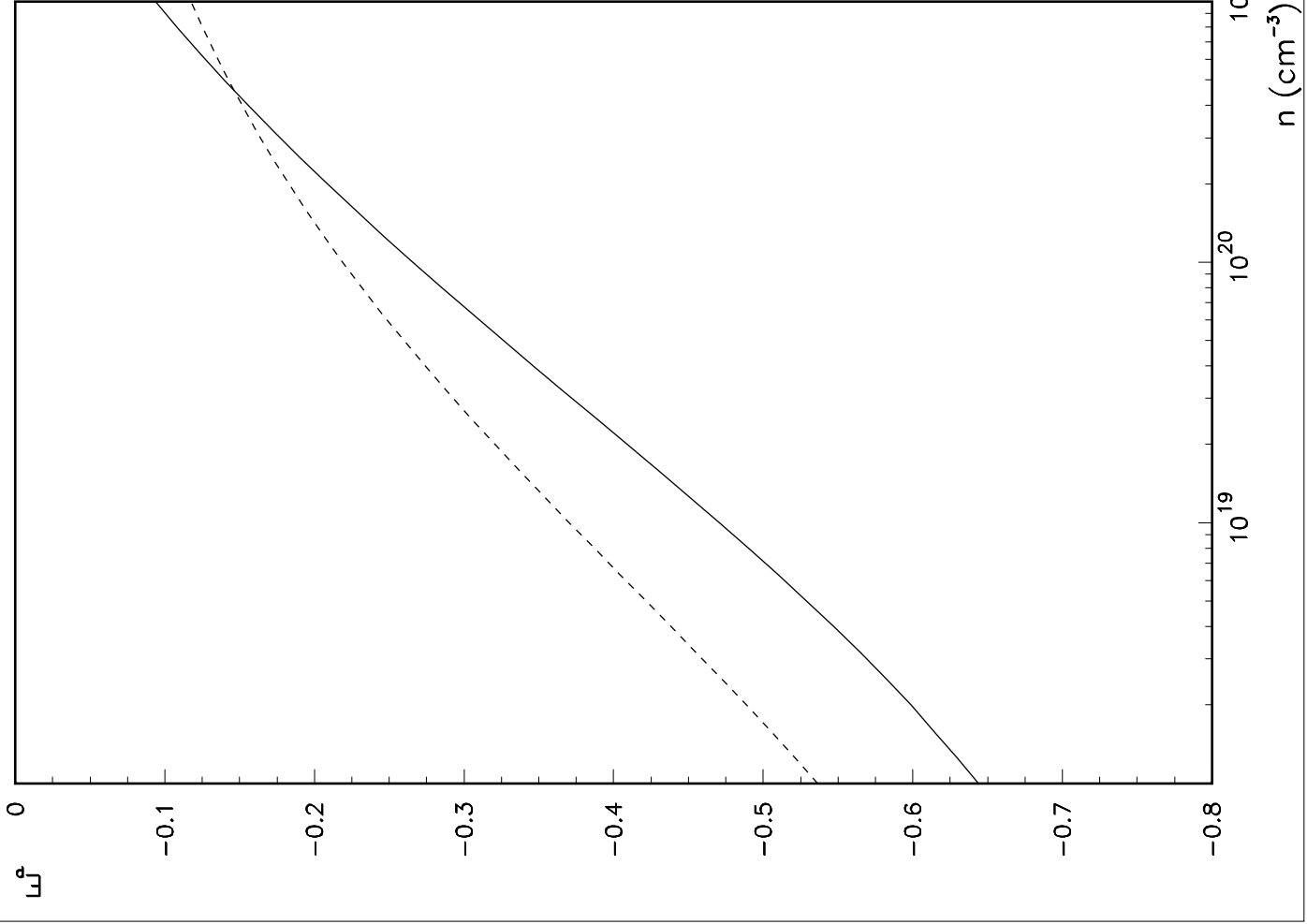
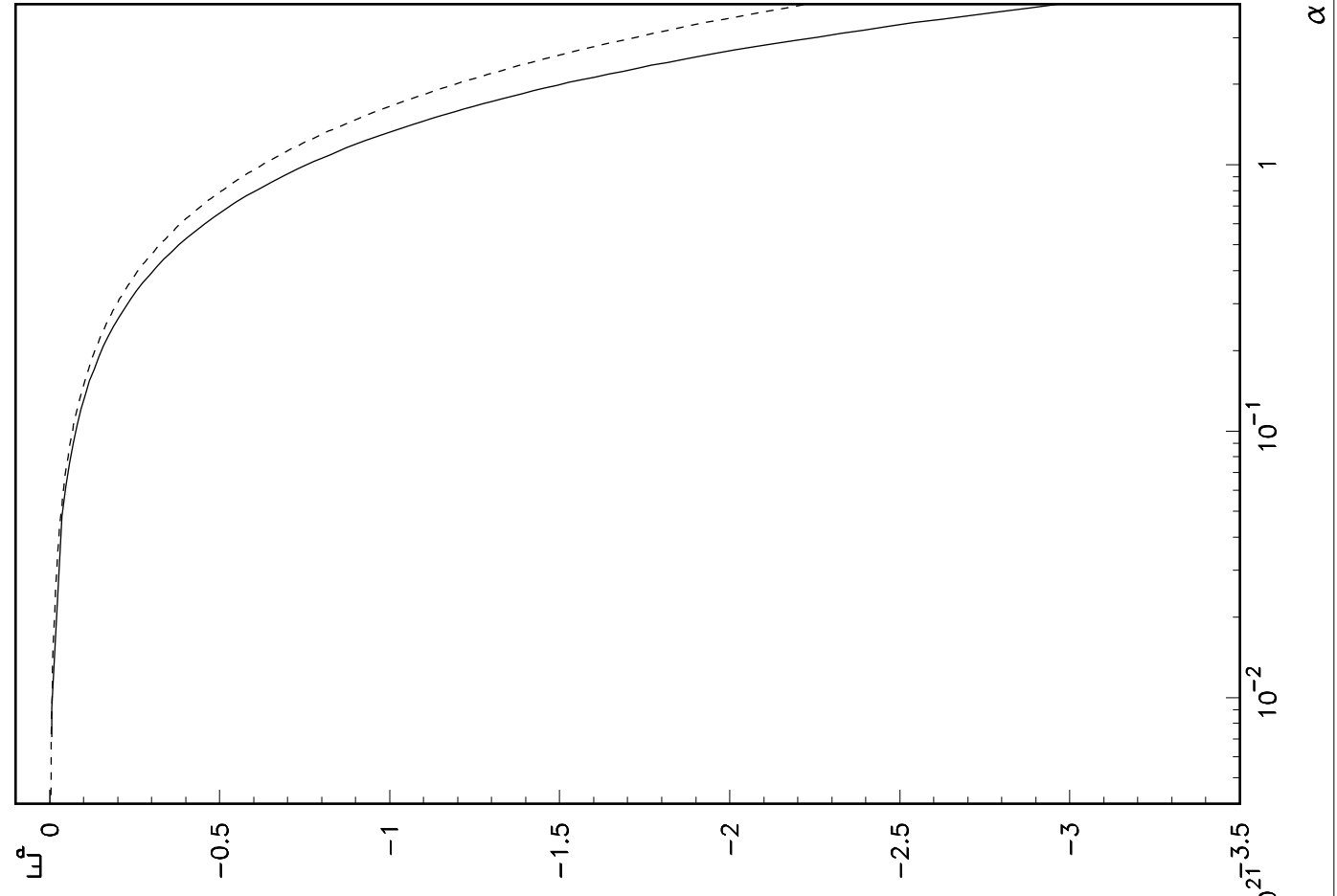
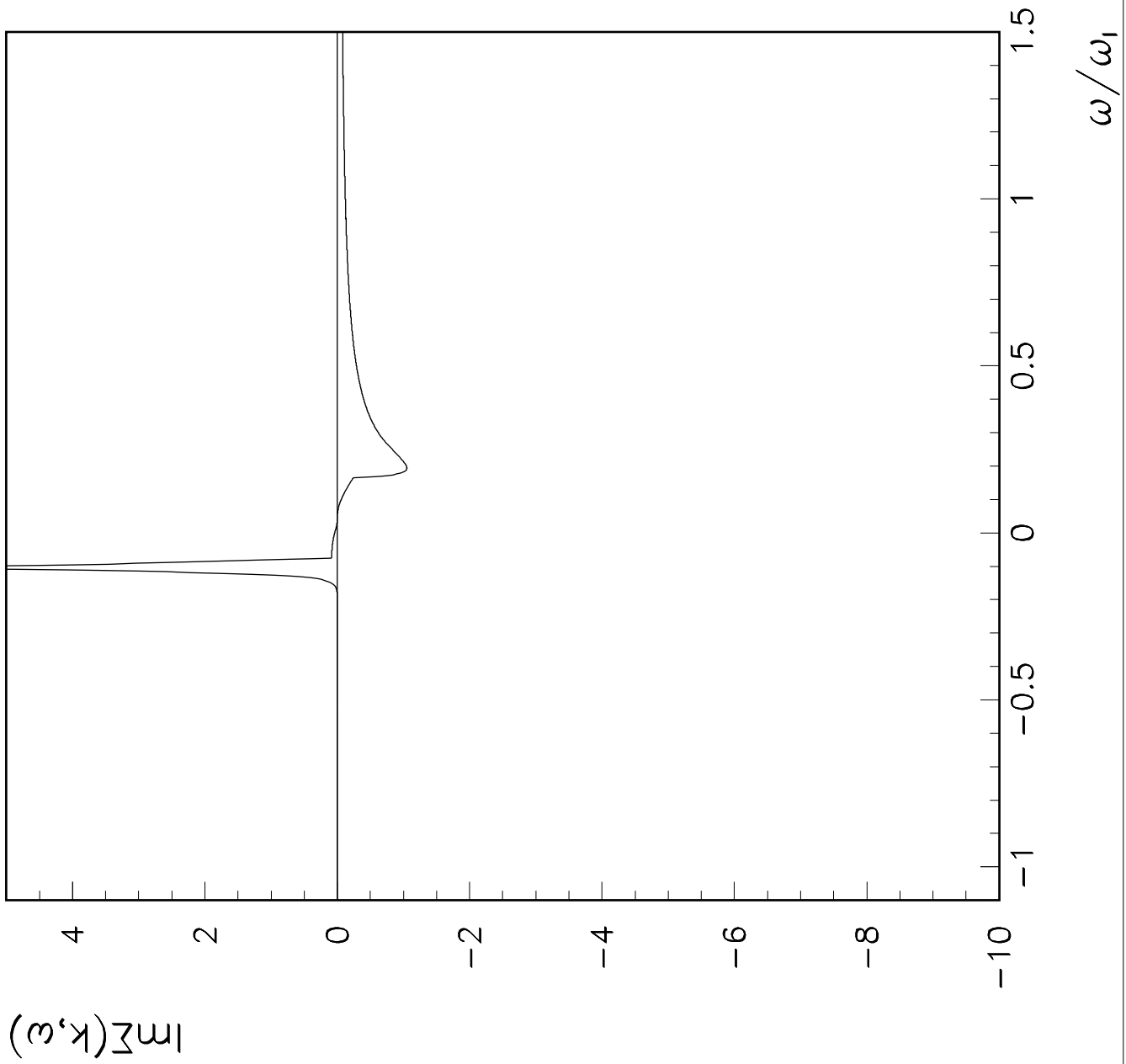
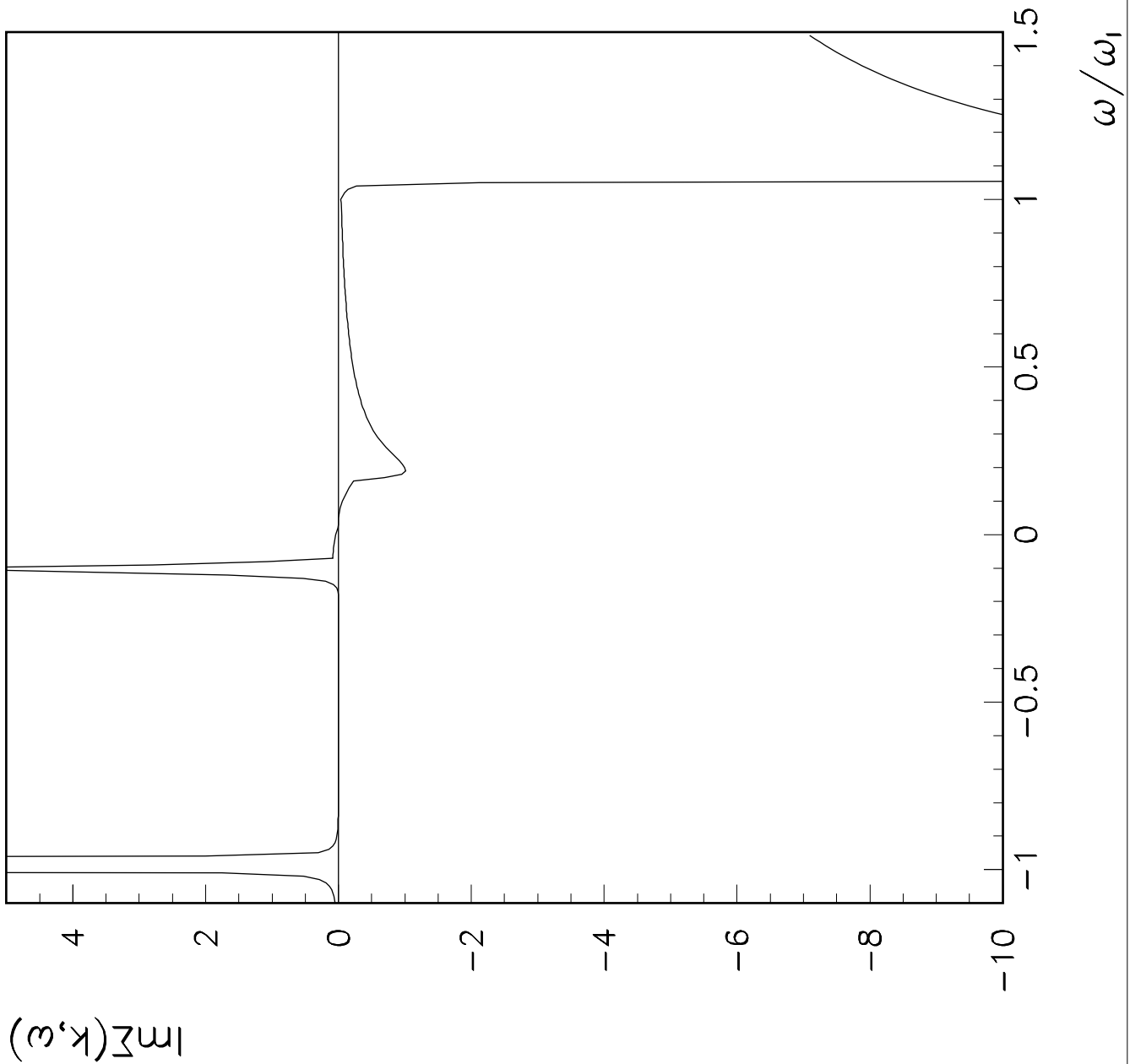
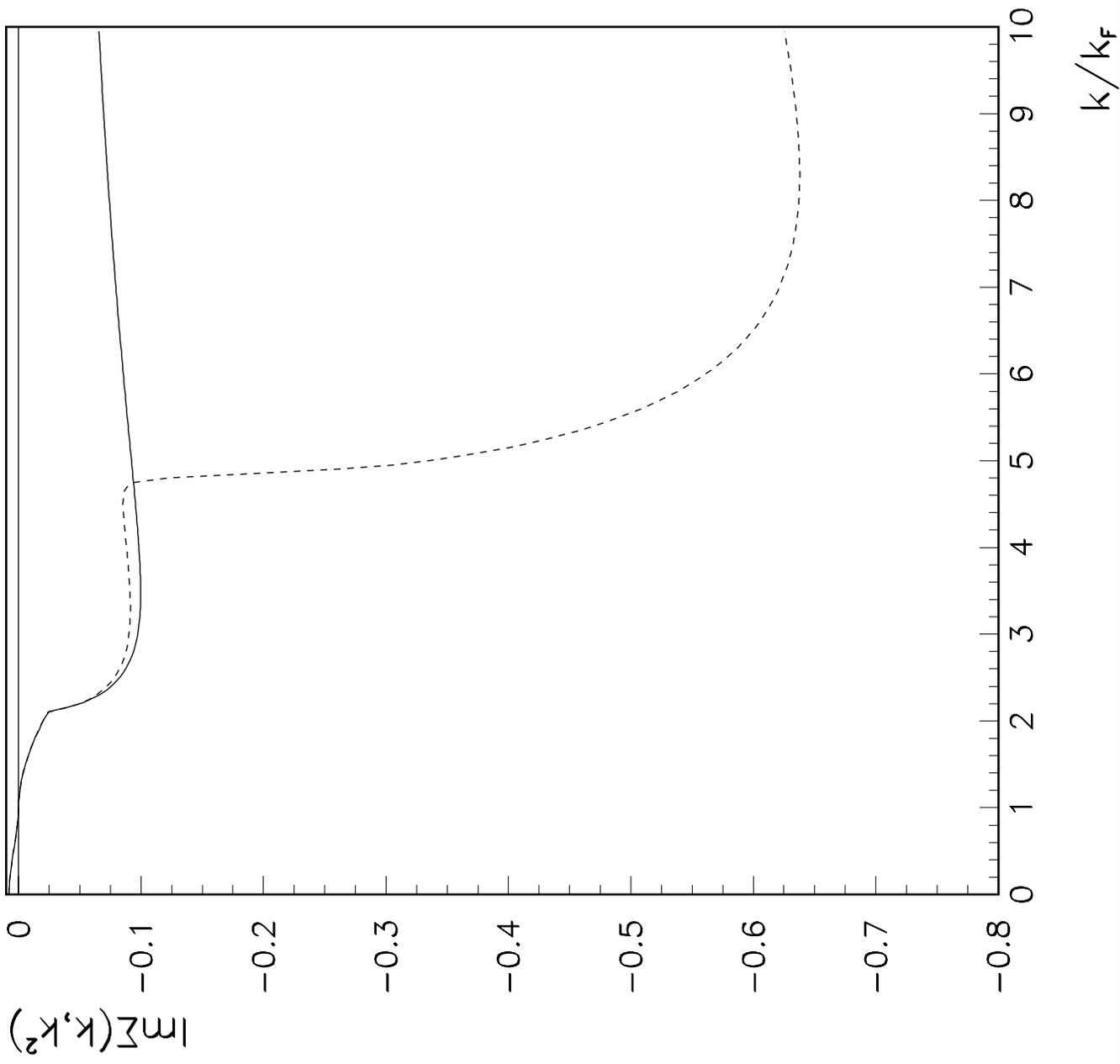
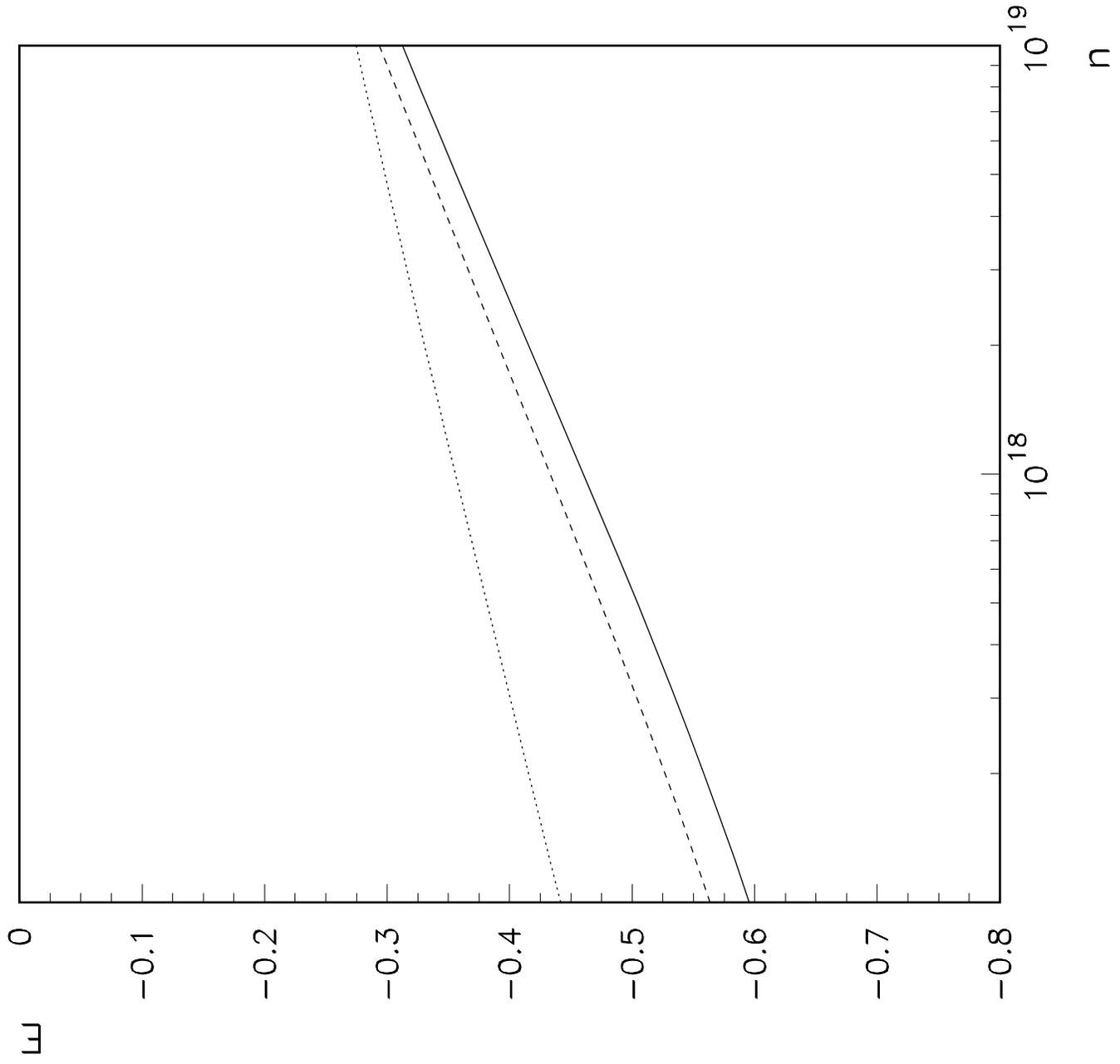
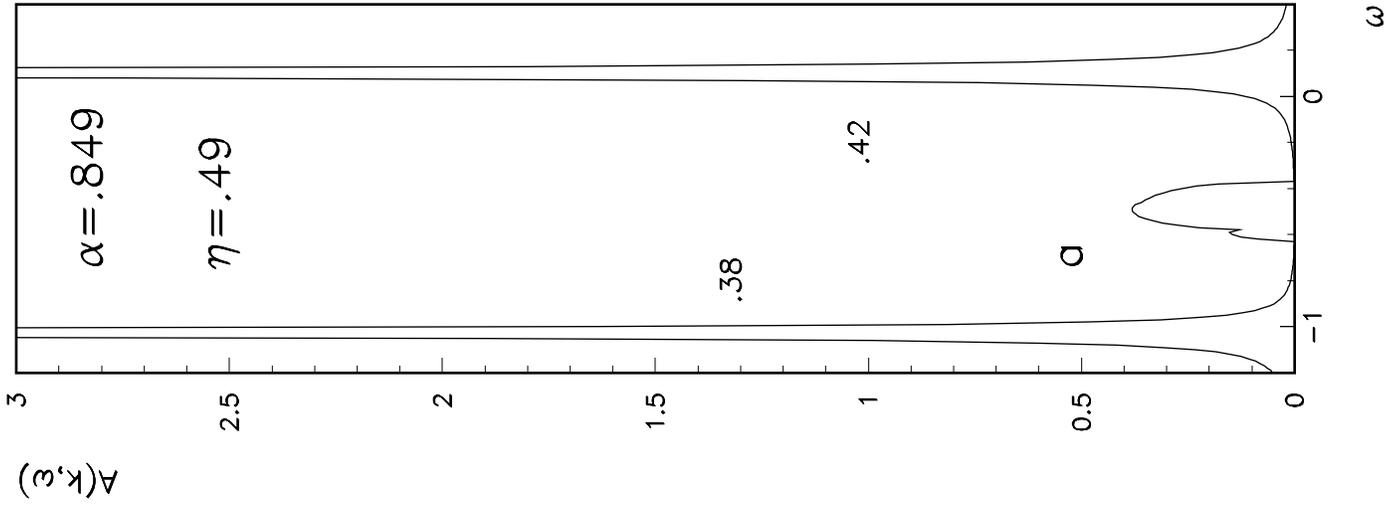
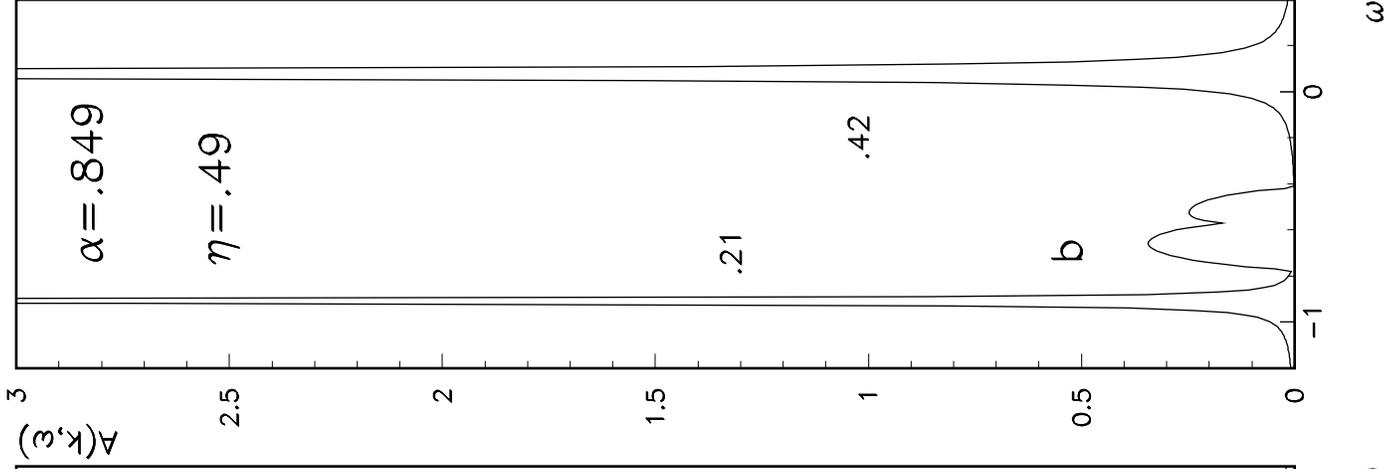
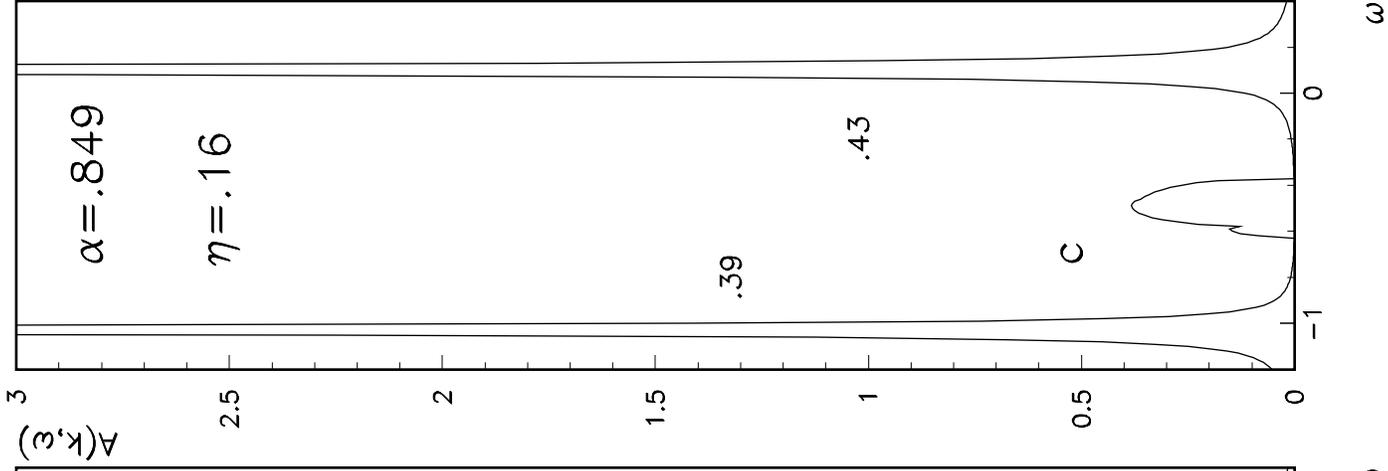
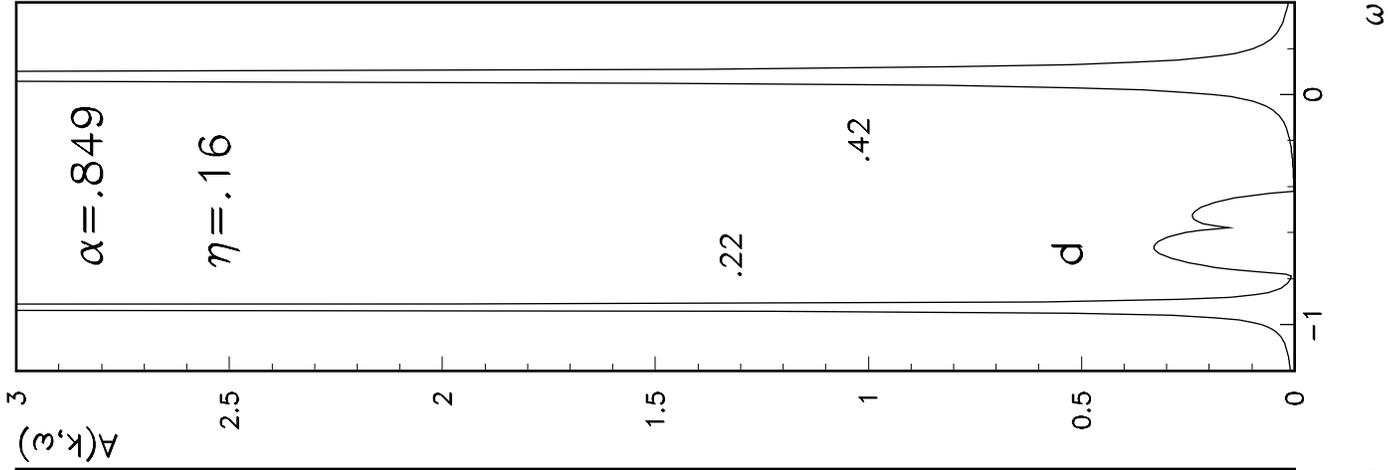
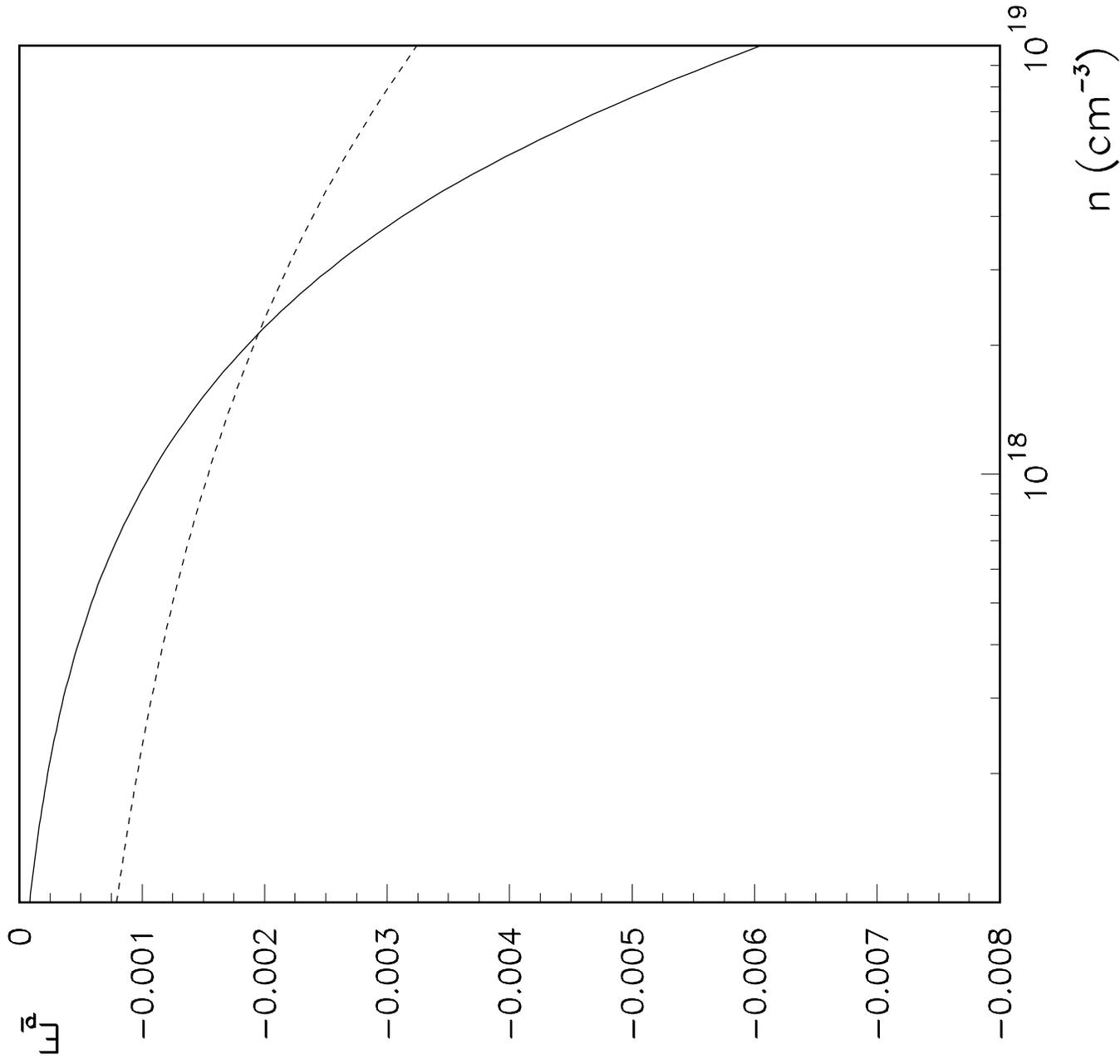
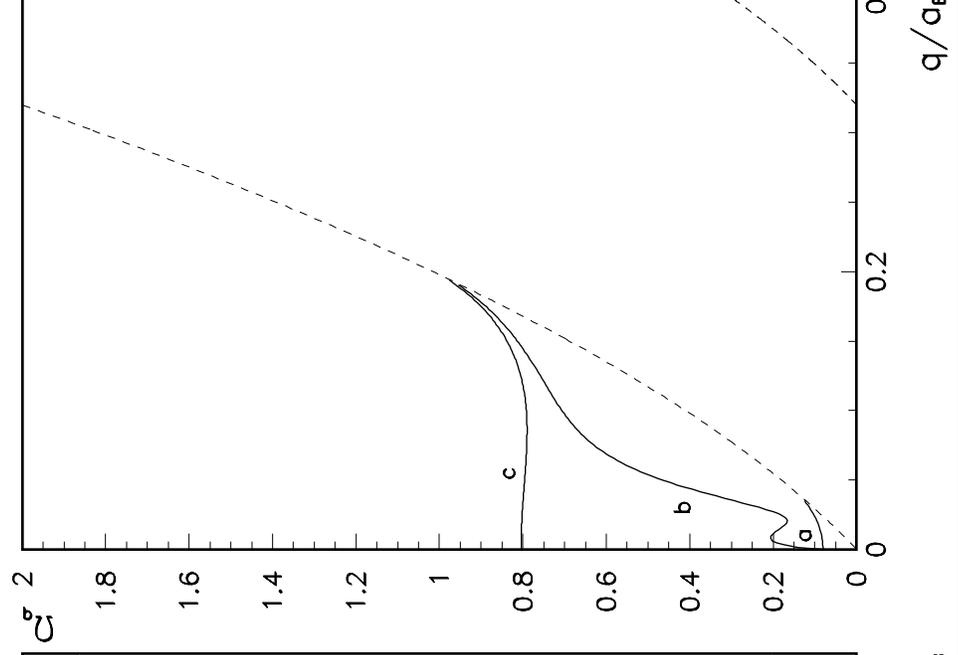
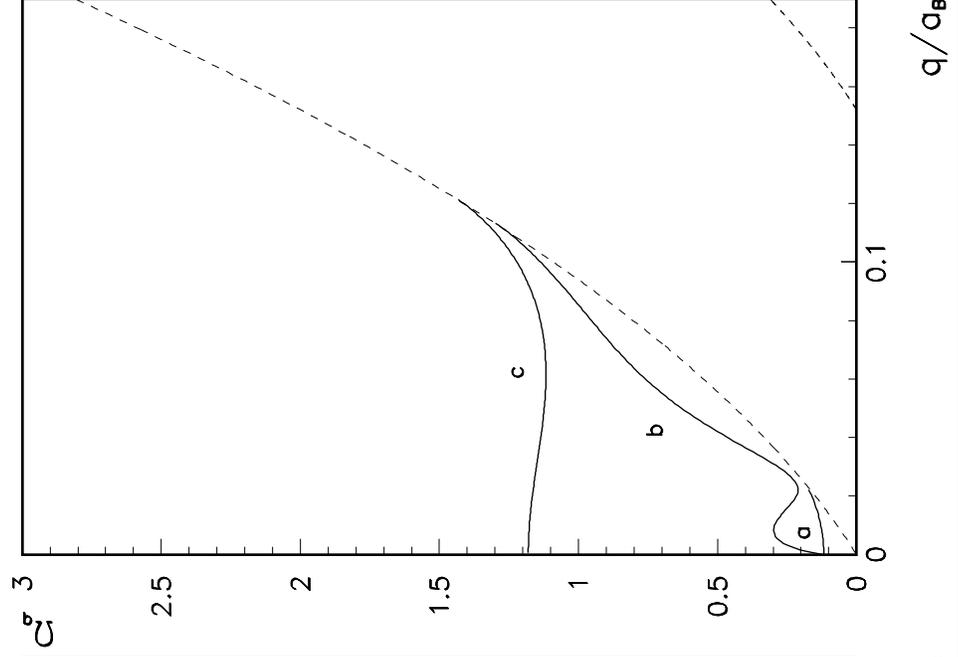
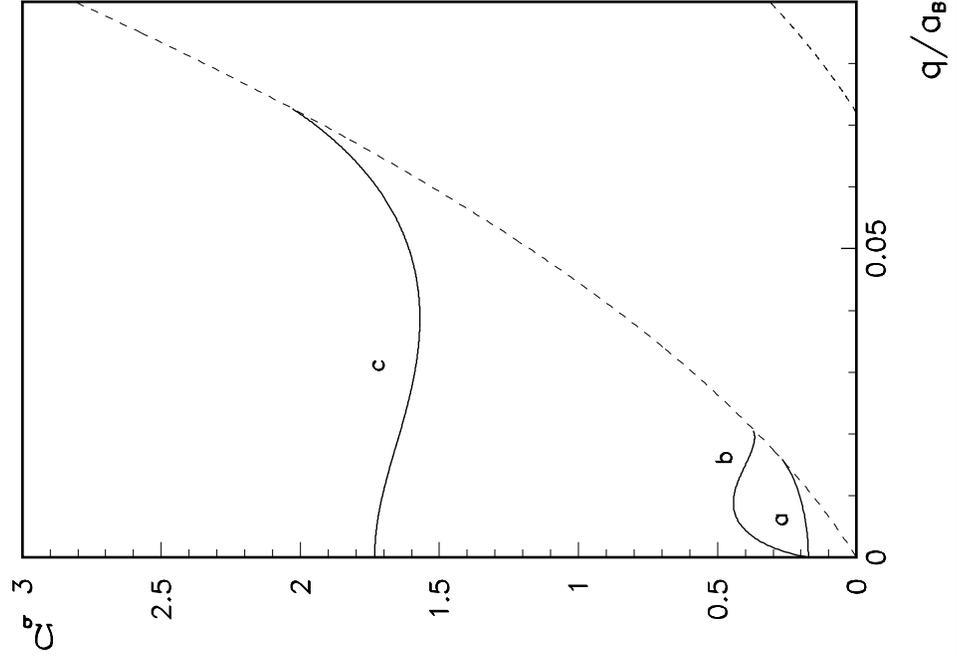
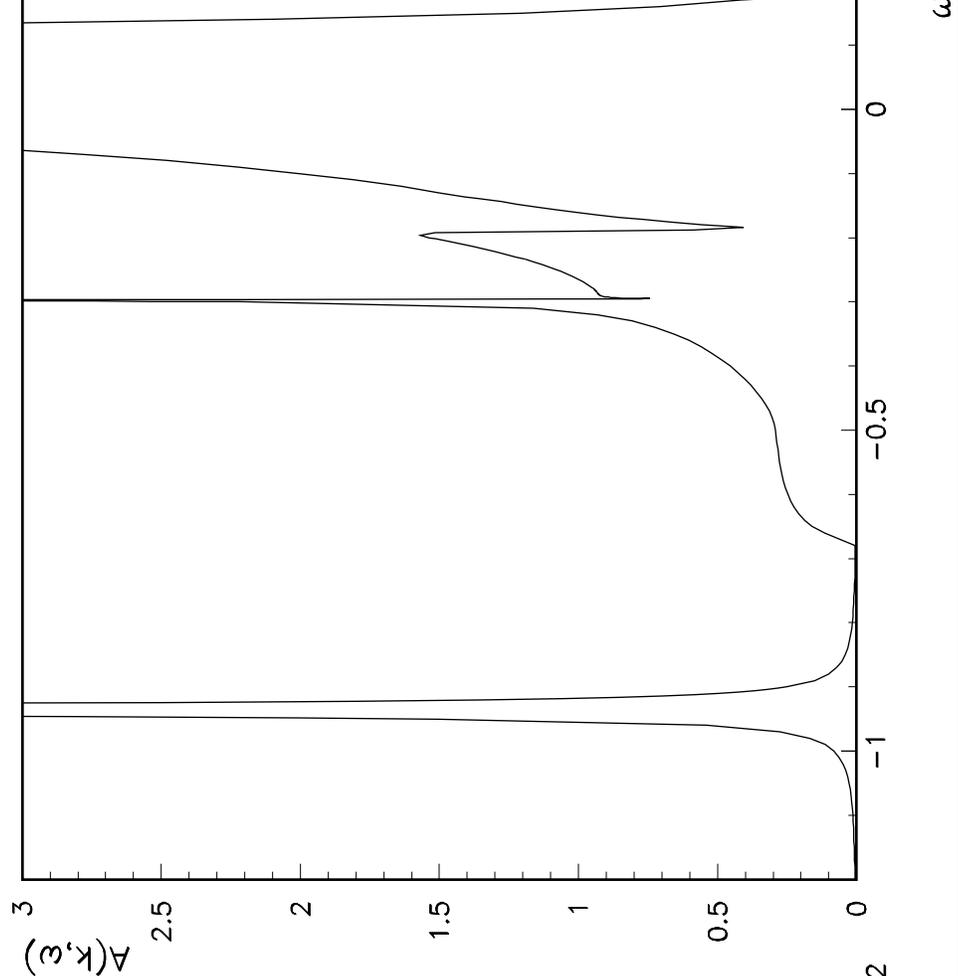
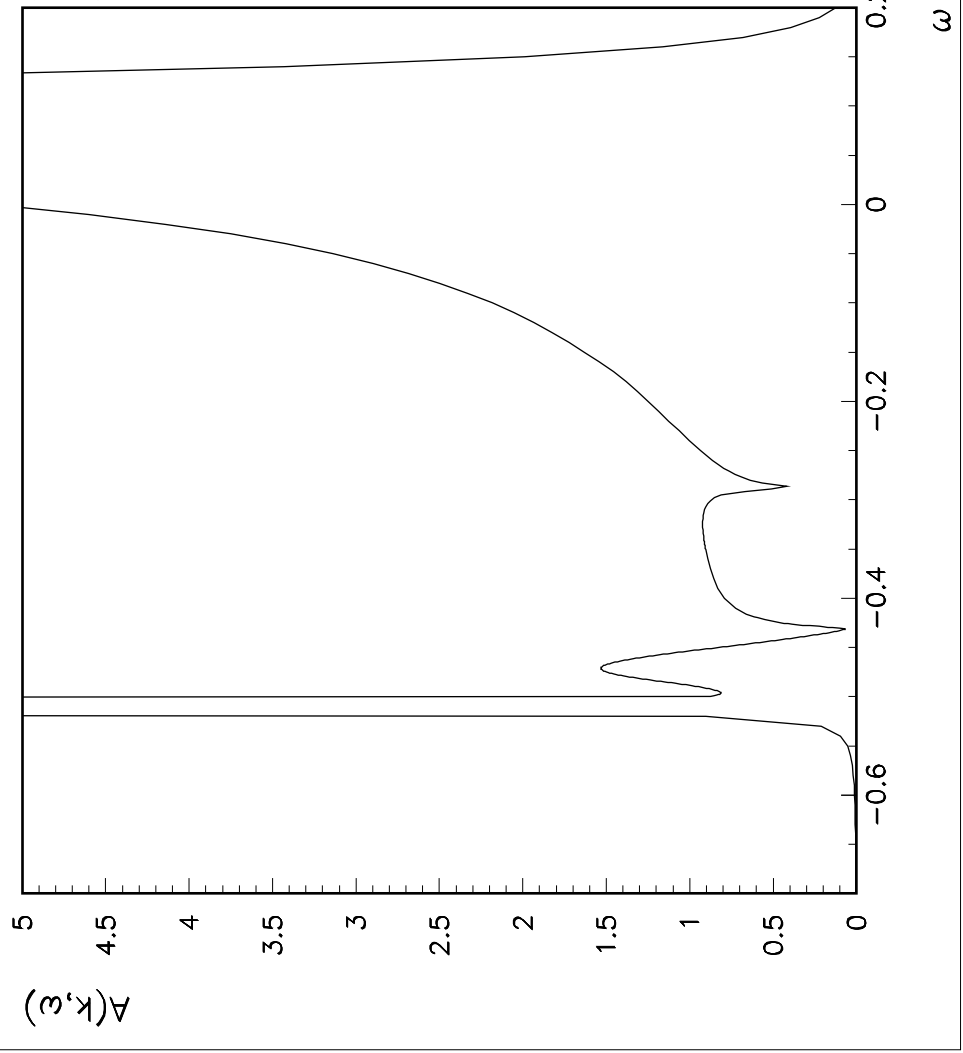
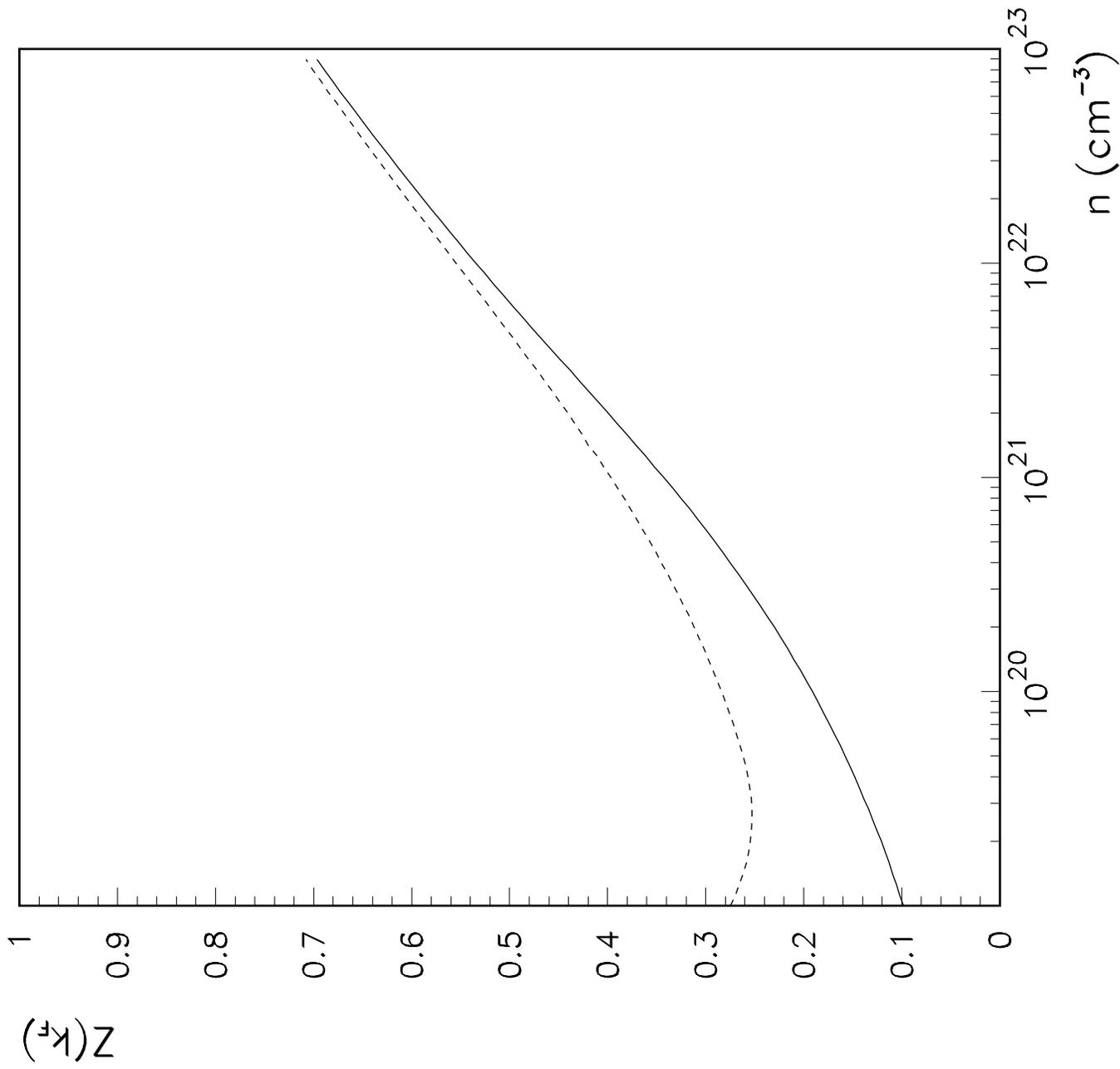
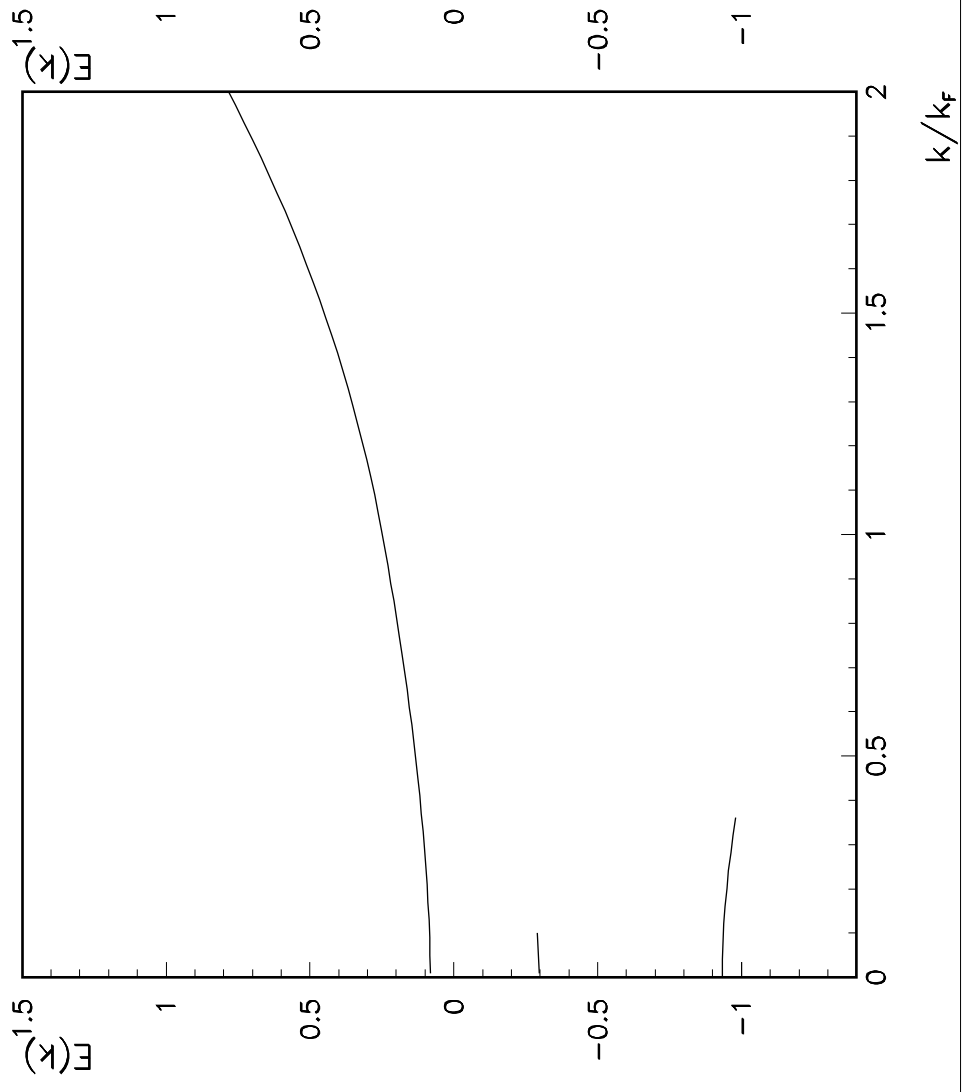
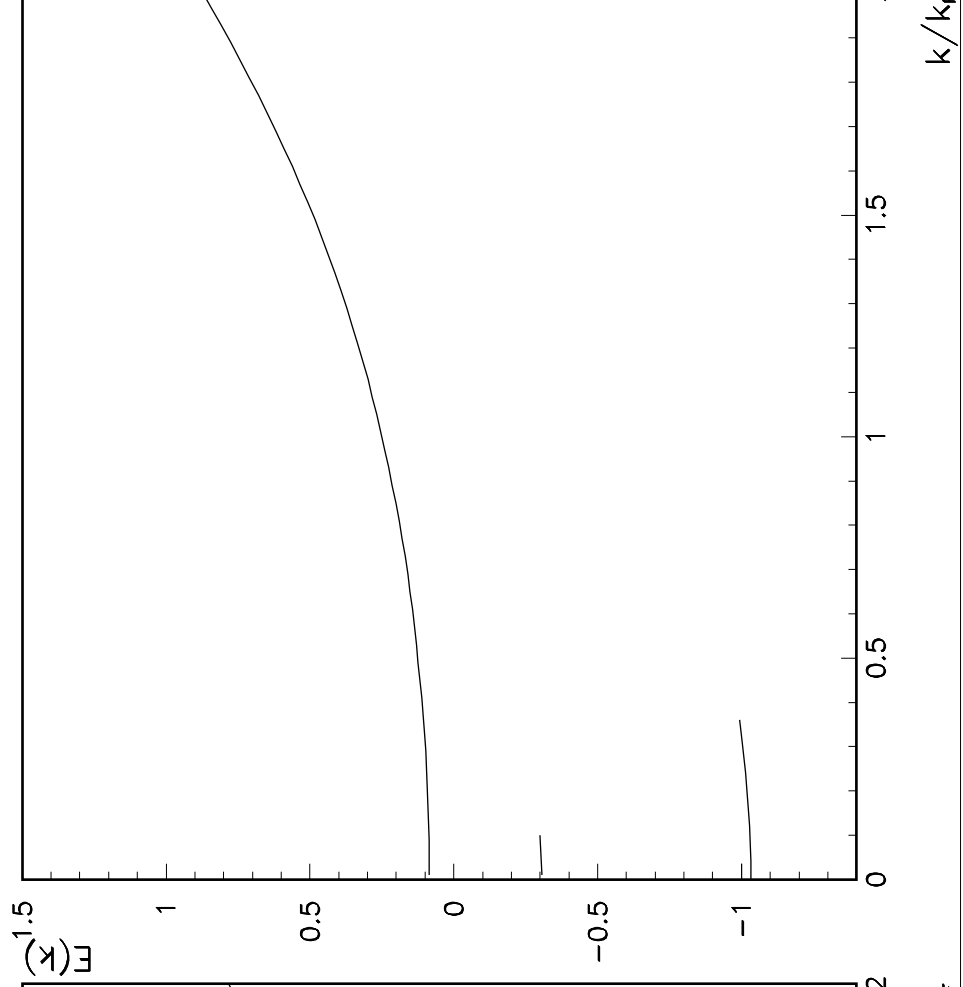
